\documentclass[journal]{IEEEtran}

\usepackage[colorlinks,urlcolor=blue,linkcolor=blue,citecolor=blue]{hyperref}
\usepackage{amsmath,amssymb,amsfonts,amsthm}
\usepackage{tabularx,booktabs, multirow}
\usepackage{graphicx}
\usepackage{bm}
\usepackage{algpseudocode}
\usepackage{algorithm}
\usepackage{physics}
\usepackage[caption=false]{subfig} 
\theoremstyle{definition}

\newtheorem{definitionx}{Definition}
\newtheorem{theoremx}{Theorem}
\newtheorem{lemmax}{Lemma}

\newtheorem{remarkx}{Remark}
\newtheorem{propertyx}{Property}
\newtheorem{propositionx}{Proposition}
\DeclareMathOperator*{\argmin}{arg\,min}

\newcommand{\tp}{^\top}
\newcommand{\Fro}{_\mathrm{F}}
\newcommand\normx[2]{\Big\lVert #1 \Big\rVert} 
\DeclareDocumentCommand\diag{}{\opbraces{\operatorname{diag}}}

\DeclareDocumentCommand\tr{}{\opbraces{\operatorname{tr}}}
\DeclareDocumentCommand\vec{}{\opbraces{\operatorname{vec}}}
\DeclareDocumentCommand\cov{}{\opbraces{\operatorname{cov}}}

\begin{document}


\title{Geometry-Aware Edge-State Tracking for \\Resilient Affine Formation Control}


\author{Zhonggang Li,~\IEEEmembership{Graduate Student Member,~IEEE,}
        Raj Thilak Rajan,~\IEEEmembership{Senior Member, IEEE}
\thanks{This work is funded by the Sensor AI Lab, under the AI Labs program of Delft University of Technology. The authors are with Signal Processing Systems, Faculty of Electrical Engineering, Mathematics and Computer Science (EEMCS), Delft University of Technology, Delft,
The Netherlands. \{z.li-22, r.t.rajan@tudelft.nl\}.}}

\maketitle

\begin{abstract}
Affine formation control (AFC) is a subset of formation control methods that enables coordinated multiagent movement while preserving affine relationships, and has recently gained increasing popularity due to its broad applicability across diverse applications. AFC is inherently distributed, where each agent's local controller relies on the relative displacements of neighboring agents. The unavailability of these measurements in practice, due to node or communication failures, leads to a change in the underlying graph topology and subsequently causes instability or sub-optimal performance. In this work, each edge in the graph is modeled using a state-space framework, allowing the corresponding edge-states to be estimated with or without up-to-date measurements. We then propose a Kalman-based estimation framework where we fuse both temporal information from agents' dynamics and spatial information, which is derived from the geometry of the affine formations. We give convergence guarantees and optimality analysis on the proposed algorithm, and numerical validations show the enhanced resilience of AFC against these topology changes in several practical scenarios.
\end{abstract}


\maketitle

\section{INTRODUCTION}
Multiagent systems have been widely researched for their broad applications in various fields such as artificial intelligence \cite{Heuillet2022MasAI}, swarm robotics \cite{Rizk2019SwarmRob}, social networks \cite{Urena2019Social}, and space applications \cite{Zhao2023Satellite}. Distributed formation control is a fundamental task in a swarm where agents collectively preserve a geometric pattern, i.e., a formation, with application to collective object transport \cite{Mora2017objtrans}, space-based interferometry \cite{Liu2018satellite}, or various sensing missions \cite{Ghamry2016forestfire}. It relies on measurements of relative information among agents to achieve different formations. For instance, distance measurements are used to achieve rigid formations that allow translations and rotations \cite{Kang2014dist, Vu2021dist}, bearing measurements allow formations with scaling of the geometry \cite{Zhao2017bearing, Zhao2022bearing}, and displacements can be employed to realize affine formations \cite{Lin2016affine}. 

Affine formation control (AFC) is a subset of displacement-based formation control, where the configurations of agents are allowed up to affine transformations over time, which include rotation, translation, shearing, or their combinations. This flexibility is favored in complex and cluttered environments where certain obstacles must be avoided. The agents can maintain a static formation \cite{Lin2016affine} or track a time-varying target configuration with continuously changing affine transformations \cite{Zhao2018affine}. This maneuverability of affine formations is typically achieved by setting a small subset of agents as leaders, which can access global and absolute information. However, rigidity conditions are imposed on the underlying graph to ensure stability, which requires careful topology design. AFC is also a case of a networked control system (NCS) \cite{zhang2019networked}, where information needs to be communicated over a network in which the communication pattern is defined by edges in a graph. Disruptions in sensing or communication, or an agent's malfunctioning in practice, will effectively change the underlying graph, which may cause stability and optimality issues. For a resilient deployment in harsh environments, strategies for topology changes are motivated. 

Solutions to topology changes in NCS are abundant in the literature, from physical layers to controller designs. The predominant strategy is to model a switched system with time-varying graphs \cite{DONG2016switching, Wang2021switching, zhai2021resilient}, where the stability is shown for every allowed switch pattern. These works are suitable for various scenarios, including time-varying edge sets \cite{DONG2016switching, Wang2021switching} where there are no missing agents, and the time-varying vertex sets \cite{zhai2021resilient} where the uncontrolled or uncooperative nodes are considered in the switching pattern. However, the AFC has a stringent graph rigidity condition to guarantee stability \cite{Lin2016affine}, which can be compromised by any missing edge or node in the predefined graph. This makes it difficult to design stabilizing patterns, especially for large-scale networks. Random and rapid changes in the graph pose additional challenges in the switched system. Another line of work uses local predictors to estimate missing information originating from topology changes, such as packet dropouts \cite{Franze2018dropout, Gong2013dropout}. The philosophy is to design only one controller for a predefined graph and develop an additional observer or predictor for the control feedback. Examples of predictor design include Kalman filtering based on dynamic models \cite{lee2018relative}, neural network-based prediction \cite{Leila2022packetloss}, etc. These approaches are not sensitive to the graph stability conditions, but often rely on the availability of measurements to prevent error accumulations, i.e., permanent information loss, such as a departing or uncooperative node, is not tolerated. In the context of AFC, we have seen very little work to address the topological changes due to the above challenges.

In this paper, we develop a distributed estimation framework for topology changes due to missing information for the affine formation maneuver control of multiagent systems. We adopt a similar philosophy to \cite{Martijn2022KF} where each edge in the graph, which denotes displacements used for local controllers, is modeled by a state-space tracking system. Then, a distributed Kalman filter can naturally be derived to estimate the states with or without edge measurements, which we use as a baseline approach. As mentioned, this local prediction works for intermittent measurements but would collapse in scenarios with failing nodes that provide no edge measurements. However, we explore alternative estimators from the geometry information embedded in the formation itself, which are then adaptively used as extra observations to the state-space model. We thus enable a unifying approach for several adverse scenarios challenging for conventional methods in the context of AFC. Compared with our preliminary results in \cite{Li2023GARKF}, additional contributions in this paper include: 1) we enhance the feasibility for the geometric estimators in more extreme conditions using a dynamic consensus filter; 2) we mathematically unveil the convergence properties for the adaptive element, known as the convergence indicator, which indicates the quality of the geometric estimates; 3) the convergence and optimality of the proposed algorithm is shown; 4) additional simulation is added to verify the claimed properties of our algorithms. 

\textit{Layout:} Section \ref{sec: preliminaries} provides the basics of the affine formation control, followed by the proposed adaptive edge tracking modeling in Section \ref{sec: prob form}. In Section \ref{sec: est design}, we introduce the way to calculate edge estimates from the formation geometry and discuss the limitations and solutions. We then propose the geometry-aware relative Kalman filtering (GA-RKF), which adaptively combines the geometric edge estimates, direct edge measurements, and dynamic predictions. The convergence and optimality of GA-RKF are discussed in Section \ref{sec: cvg and opt} before numerical validations of the proposed concepts and practical scenarios in Section \ref{sec: sim}. Finally, we provide conclusions and future works in Section \ref{sec: conclusions}.

\textit{Notations:} Vectors and matrices are represented by lowercase and uppercase boldface letters, respectively, such as $\mathbf{a}$ and $\mathbf{A}$. The elements of matrix $\mathbf{A}$ are expressed using $[\mathbf{A}]_{ij}$ where $i$ and $j$ denote the index of rows and columns, respectively. Sets and graphs are represented using calligraphic letters, e.g., $\mathcal{A}$. The relative complement of two sets is denoted by \textbackslash.  Vectors of length $N$ of all ones and zeros are denoted by $\mathbf{1}_N$ and $\mathbf{0}_N$, respectively. An identity matrix of size $N$ is denoted by $\mathbf{I}_N$. The Kronecker product is $\otimes$ and a vectorization of a matrix is denoted by $\vec(\cdot)$ by stacking all the columns vertically. The trace operator is denoted by $\tr(\cdot)$. We also use subscripts $k$ on vectors and matrices to indicate a time-varying variable.

\section{PRELIMINARIES}\label{sec: preliminaries}
\subsection{Graph Theory} We consider $N$ mobile agents in a $D$-dimensional space with $D$ typically being $2$ or $3$. The prescribed interactions among the agents are described by an undirected \textit{nominal graph} $\mathcal{G}=(\mathcal{V},\mathcal{E})$ where the set of vertices is $\mathcal{V}=\{1,...,N\}$ and the set of edges is $\mathcal{E}\subseteq\mathcal{V}\times\mathcal{V}$. As an undirected graph is equivalent to a bidirectional graph, i.e., $(i,j)\in\mathcal{E} \Leftrightarrow (j,i)\in\mathcal{E}$, we assume there are $M$ directed edges (or $M/2$ undirected edges) in total. The set of neighbors of a node $i\in\mathcal{V}$ is defined as $\mathcal{N}_i = \{j\in\mathcal{V}: (i,j)\in\mathcal{E}\}$, and the corresponding cardinality is denoted by $N_i$ with $\sum_{i\in\mathcal{V}}N_i=M$. In operation, the time-varying connectivity of the system is modeled by a \textit{functional graph} $\mathcal{G}_k=(\mathcal{V}_k,\mathcal{E}_k)$ where $k$ is the discrete-time index. The functional graph is a subgraph of the nominal graph for all time instances, i.e., $\mathcal{V}_k\subseteq\mathcal{V}$ and $\mathcal{E}_k\subseteq\mathcal{E}$ for all $k$. There are $N_k \leq N$ nodes and $M_k \leq N$ directed edges in the functional graph at time index $k$, and the time-varying cardinality of the set of neighbors is denoted by $N_{i,k}$ with $N_{i,k}\leq N_{i}$ which satisfy $\sum_{i\in\mathcal{V}_k}N_{i,k}=M_k$. The difference between functional and nominal graphs represents the failure of nodes and edges.

\begin{figure}[t]%
	\centering%
	\includegraphics[width=0.83\linewidth]{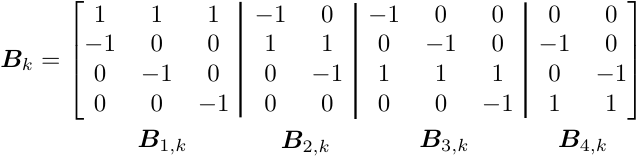}%
	\caption{An example of the incidence matrix $\bm{B}_k$, which implies 10 (directed) edges. Note that this graph has fewer edges compared to its fully connected nominal counterpart.}%
	\label{fig: inc mat}%
\end{figure}

In this work, a frequently used graph characterization is the incidence matrix, and its definition for the functional graph $\bm{B}_k\in\mathbb{R}^{N\times M_k}$ is defined as
\begin{equation}\label{equ: def inc mat}
    [\bm{B}_k]_{im}=\begin{cases}
  1& \text{if edge } e_m = (i,j) \in\mathcal{E}_k \text{ leaves node } i \\
  -1& \text{if edge } e_m = (i,j)\in\mathcal{E}_k \text{ enters node } i \\
  0&  \text{otherwise}
\end{cases},
\end{equation}
which admits the structure $\bm{B}_k = [\bm{B}_{1,k},\bm{B}_{2,k},...,\bm{B}_{N,k}]$ where the matrix $\bm{B}_{i,k}\in\mathbb{R}^{N\times N_i},\forall i \in \mathcal{V}$, groups the columns per agent. Note that $\bm{1}_N\tp\bm{B}_{i,k}=\bm{0}_{N_{i,k}}\tp$. Fig. \ref{fig: inc mat} shows an example of $\bm{B}_k$ for a functional graph rooted from a 4-node complete nominal graph.

\subsection{Geometric Transformations}
We define $\bm{z}_{i,k}\in\mathbb{R}^{D}$ as the time-varying position of agent $i$ at time instant $k$, and collect all the positions of the agents in a \textit{configuration matrix} $\bm{Z}_k=[\bm{z}_{1,k},\bm{z}_{2,k},...,\bm{z}_{N,k}]\in\mathbb{R}^{D\times N}$. Similarly, a \textit{target configuration} that collects the expected positions of agents is defined as $\bm{Z}_k^*=[\bm{z}_{1,k}^*,\bm{z}_{2,k}^*,...,\bm{z}_{N,k}^*]\in\mathbb{R}^{D\times N}$ in which $\bm{z}_{i,k}^*\in\mathbb{R}^{D}$ is the individual target position for each agent $i\in\mathcal{V}$. For the nominal graph $\mathcal{G}$, we introduce a \textit{nominal configuration} $\bm{P}=[\bm{p}_{1},\bm{p}_{2},...,\bm{p}_{N}]\in\mathbb{R}^{D\times N}$ where $\bm{p}_i\in\mathbb{R}^{D}$ is the nominal position for agent $i$. The nominal configuration represents a general geometric pattern that the agents are expected to achieve, and the time-varying target configuration is a time-varying mapping from the nominal configuration through geometric transformations.

\begin{figure}[t]
	\centering	
	\subfloat[nominal]{\raisebox{0ex}
		{\includegraphics[width=0.1\textwidth]{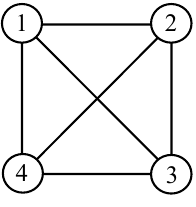}}%
	}
	\hspace{1ex}
	\subfloat[scaling]{\raisebox{0ex}
		{\includegraphics[width=0.08\textwidth]{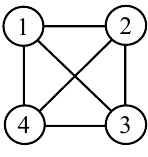}}%
	} 
	\hspace{1ex}
	\subfloat[rotation]{\raisebox{0ex}
		{\includegraphics[width=0.12\textwidth]{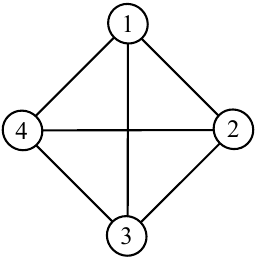}}%
	}
	\hspace{1ex}
	\subfloat[shearing]{\raisebox{0ex}
	{\includegraphics[width=0.13\textwidth]{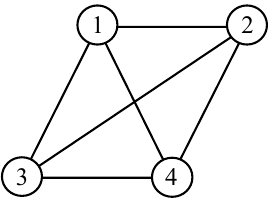}}}%

	\caption{An illustration of a few basic affine transformations.}
	\label{fig: affine trans example}
\end{figure}

In the context of the affine formation control (AFC), the transformation is limited to affine transformations, which pose the following constraint 
\begin{equation}\label{equ: target config}
    \bm{Z}^*_k = \bm{\Theta}^*_k\bm{P} + \bm{t}^*_k\bm{1}_N\tp,
\end{equation}
where $\bm{\Theta}_k^*\in\mathbb{R}^{D\times D}$ characterizes a shape transformation such as scaling, rotation, shearing, etc., and $\bm{t}_k^*\in\mathbb{R}^{D}$ indicates a collective translation. Some illustrations of the affine shape transformation are shown in Fig. \ref{fig: affine trans example}, and note that they pose constraints on the structure of the $\bm{\Theta}_k^*$ matrix with a reduced degree of freedom. In summary, the continuous target positions, which form several trajectories, are generated by designing $\bm{\Theta}_k^*$ and $\bm{t}_k^*$, given the nominal positions of the agents. The goal for the formation maneuver control is to steer the agents such that their positions $\bm{Z}_k$ converge to and track the target $\bm{Z}_k^*$ with minimum error.

\subsection{Leader-Follower Maneuver Control}\label{sec: afc}
AFC is a consensus-based control that aims to steer the agents into their desired positions using relative information, i.e., displacements. Typically, a small subset of agents $\mathcal{V}_l\subseteq\mathcal{V}$ is set as \textit{leaders} that globally track the target positions while still interacting with the remaining majority, i.e., the set of \textit{followers} $\mathcal{V}_f=\mathcal{V}\setminus \mathcal{V}_l$. In this scheme, target configurations $\bm{Z}_k^*$ are achieved but not informed to the followers if they maintain the formation. In other words, the geometry patterns for the maneuvers are described by the leaders. To achieve AFC, the number of leaders required is $D+1$ with some geometrical conditions and typically $N \gg D+1$ \cite{Zhao2018affine}. Hence, we limit our discussions to the followers in terms of control laws, edge and node failures, and the prospective estimator designs in this paper.


In this work, we consider single-integrator dynamics in discrete time for the agents i.e., $\forall i \in\mathcal{V}_f$ we have $\bm{z}_{i,k+1} = \bm{z}_{i,k} + \Delta t\bm{u}_{i,k}$, where $\Delta t$ is a small time interval and $\bm{u}_{i,k}$ is the velocity input. The control laws for various scenarios are shown in  Table \ref{tab: control laws} \cite{Zhao2018affine,xu2020affine}. Here, we use a shorthand notation for the displacements i.e., $\bm{z}_{ij,k}=\bm{z}_{i,k}-\bm{z}_{j,k}$ and the edge weights $l_{ij}, \forall (i,j)\in\mathcal{E}$ are associated with a nominal graph. The rigidity of the nominal graph guarantees the existence of stabilizing weights \cite{Lin2016affine}, which can be calculated through convex programs \cite{Lin2016affine} or mixed-integer programs \cite{xiao2022framework} given the graph design. We initially assume the nominal graph satisfies the rigidity constraints, ensuring the stability of the control laws in Table \ref{tab: control laws}, and later focus on realistic scenarios where such rigidity conditions may not hold. 

\begin{table}[t]
	\caption{Discrete-time control laws for AFC \cite{Zhao2018affine,xu2020affine}}
	\label{tab: control laws}
	\centering
	\begin{tabular}{l l}
		\toprule
		{Leaders} & Control law\\
		\midrule
		{static} & $\displaystyle\bm{u}_{i,k} = -\sum_{j\in\mathcal{N}_i}l_{ij}\bm{z}_{ij,k}$ \\
		
		{constant velocity} & $\begin{aligned}  \bm{u}_{i,k} = -\alpha\sum_{j\in\mathcal{N}_i}l_{ij}\bm{z}_{ij,k}-\eta\sum_{\tau=0}^{k} 
\sum_{j\in\mathcal{N}_i}l_{ij}\bm{z}_{ij,\tau}\end{aligned}$   \\
		
		{varying velocity} & $\begin{aligned}  \bm{u}_{i,k}  = -\frac{1}{\gamma_i}\sum_{j\in\mathcal{N}_i}l_{ij}(\bm{z}_{ij,k}-\dot{\bm{z}}_{j,k}) \\ 
\end{aligned}$ \\	
		\bottomrule
	\end{tabular}
\end{table}
\begin{figure*}[ht]
	\centering	
	\subfloat[nominal graph]{\raisebox{0ex}
		{\includegraphics[width=0.15\textwidth]{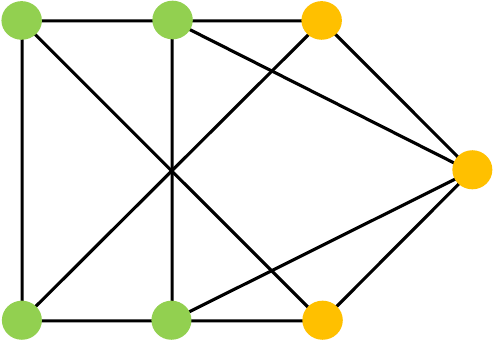}}%
	}
	\hspace{5ex}
        \subfloat[S1: directed edges]{\raisebox{0ex}
		{\includegraphics[width=0.15\textwidth]{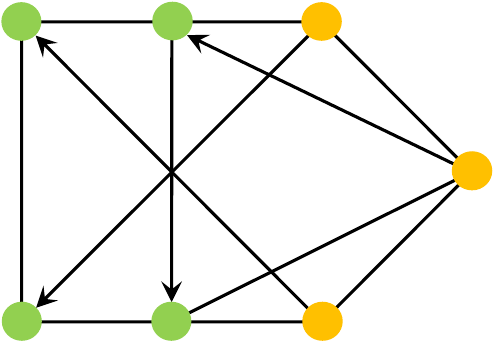}}%
	}
	\hspace{5ex}
	\subfloat[S2: not rigid]{\raisebox{0ex}
		{\includegraphics[width=0.15\textwidth]{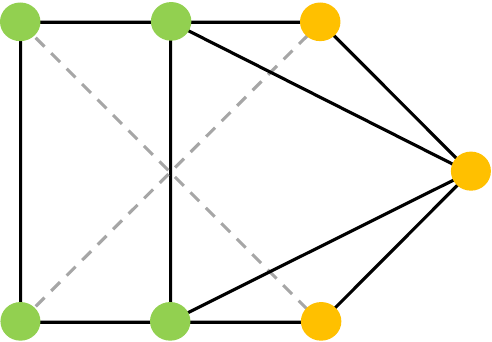}}%
	} 
        \hspace{5ex}
	\subfloat[S3: node departure]{\raisebox{0ex}
		{\includegraphics[width=0.15\textwidth]{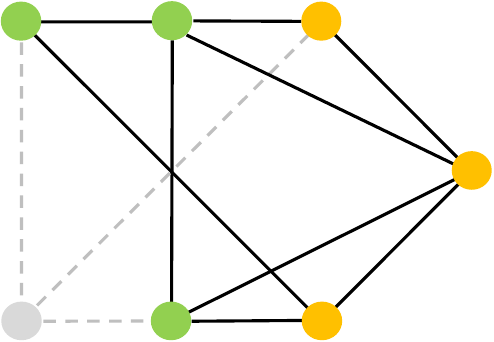}}%
	}
	\hspace{0ex}
	
	\caption{Some practical scenarios modeled by functional graphs, which are subgraphs of a nominal graph. The colored nodes and solid edges represent true agents and connections; orange nodes are candidates for leaders; gray nodes and dashed edges are unavailable agents and connections in the functional graph. The edges with arrows are directed edges representing one-way flow of information.}
	\label{fig: examples functional graph}
		
\end{figure*}

\begin{definitionx}[\textit{Sufficient formation convergence}]\label{def: formation convergence}
    We claim that the agents have sufficiently converged to their target positions at a given time instance $k$ if  
\begin{equation}\label{equ: def tracking error}
    \delta_k = \frac{1}{N}\left \| \bm{Z}_k-\bm{Z}^*_k  \right \| ^2_\mathrm{F}\leq\bar{\delta},
\end{equation}
with an arbitrarily small $\bar{\delta}$.
\end{definitionx}

\noindent In AFC frameworks \cite{Zhao2018affine, Zu2018affineDir, Su2023bearing, Xiao2022optimize}, the tracking error $\delta_k$ is a common performance metric, which globally indicates the convergence of the algorithm. Sufficient convergence is asymptotically satisfied for controllers in Table \ref{tab: control laws} with no or bounded measurement noise, but can be violated at formation initialization or under large environmental disturbances before re-stabilization.


\section{PROBLEM FORMULATION}
\label{sec: prob form}
\subsection{Motivation for Edge-State Estimation}\label{sec: scenario description}
In the previous section, we introduced formation controllers associated with a time-invariant nominal graph with certain rigidity conditions. In reality, these conditions are often not satisfied due to various practical challenges, e.g., communication constraints, sensor scheduling, or hardware failure, which result in a time-varying functional graph. A few scenarios are illustrated in Fig. \ref{fig: examples functional graph} and described below. 

\begin{enumerate}
    \item \textit{One-way communication links.} During operation, there could be instances when information flows only in one direction, resulting in a directed functional graph, as shown in Fig. \ref{fig: examples functional graph}(b). Several existing controllers have been designed for directed topologies \cite{Zu2018affineDir}, but these solutions are generally incapable of randomly and potentially rapidly changing topologies.
    \item \textit{Violation of stabilizing conditions.} A realization of the functional graph might not satisfy the required rigidity conditions for the existence of stabilizing controllers. See an example scenario in Fig. \ref{fig: examples functional graph}(c), in which the graph can lose rigidity after removing one or more edges.
    \item \textit{Departing and rejoining agents.} In practice, it is common for agents to depart from the swarm due to failure or battery life, as shown in Fig. \ref{fig: examples functional graph}(d). In this case, the rest of the agents are still expected to maintain the formation in the absence of missing agents. In such situations, the controllers designed for the original topology are no longer effective.
\end{enumerate}

To overcome these adverse scenarios, we design a state-space model for each edge, whose availability over time is modeled as intermittent observations. Furthermore, we explore the geometry of the formation that contains additional edge information to improve the estimator. Given this model, we could apply filtering and prediction techniques without having to redesign controllers or introducing switching systems.  

\subsection{Relative State Space Model} \label{sec: data model}
In this section, we introduce the edge-state tracking model and later develop estimators based on this model. For each edge $(i,j)\in\mathcal{E}$ in the nominal graph, we assign a state vector $\bm{\gamma}_{ij,k} = [\bm{z}_{ij,k}\tp,\dot{\bm{z}}_{ij,k}\tp,\ddot{\bm{z}}_{ij,k}\tp]\tp\in\mathbb{R}^{3D}$ which includes the displacement $\bm{z}_{ij,k}$ and the corresponding higher-order kinematics such as (relative) velocities and accelerations. We then propose a linear state-space model as follows
\begin{subequations}\label{equ: state space model}
\begin{align}
    \bm{\gamma}_{ij,k+1}&=\bm{F}\bm{\gamma}_{ij,k}+\bm{w}_{ij,k}\label{equ: ss dynamical model},\\
    \tilde{\bm{y}}_{ij,k} &= \bm{G}\bm{\gamma}_{ij,k}+\tilde{\bm{v}}_{ij,k}\label{equ: switching observation model},
\end{align}
\end{subequations}
where $\bm{F}\in\mathbb{R}^{3D\times 3D}$ is the state transition matrix in the dynamical model, and $\bm{G} = \mqty[1 &0 & 0]\otimes\bm{I}_D\in\mathbb{R}^{D\times 3D}$ is an observation matrix that selects only the observation of the displacement i.e., $\tilde{\bm{y}}_{ij,k}\in\mathbb{R}^D$. The noises $\bm{w}_{ij,k}\sim\bm{\mathcal{N}}(\bm{0}, \bm{Q}_{ij})$ and $\tilde{\bm{v}}_{ij,k}\sim\bm{\mathcal{N}}(\bm{0}, \tilde{\bm{R}}_{ij,k})$ are assumed to be zero-mean Gaussian. Since this state-space model is based on relative kinematics (upto a translation), we also refer to (\ref{equ: state space model}) as the relative state-space model.


The state transition matrix $\bm{F}$ and the noise covariance of the dynamical model $\bm{Q}_{ij}$ can be designed readily for various scenarios, e.g., constant velocity, constant acceleration, etc. \cite{bar2011tracking}. In case of the observation model (\ref{equ: switching observation model}), the state vector measurements should be incorporated, if available. However, these measurements may be unavailable due to various practical challenges discussed in the earlier section. In this case, we propose geometric estimators exploiting the affine formation properties of the network. Therefore, the observation model should choose either the available measurements or the geometric solution. The composite observation $\tilde{\bm{y}}_{ij,k}$ is then
\begin{equation}\label{equ: switching observations def}
\tilde{\bm{y}}_{ij,k} = 
\begin{cases}
\bm{y}_{ij,k},&\text{if } (i,j)\in\mathcal{E}_k \\
\hat{\bm{z}}_{ij,k}^{\text{geo}}, &\text{if } (i,j)\in{\mathcal{E}\setminus\mathcal{E}_k}
\end{cases},
\end{equation} 
where 
\begin{equation}\label{equ: obs model}
\bm{y}_{ij,k} = \bm{z}_{ij,k}+\bm{v}_{ij,k},
\end{equation}
is the observation available to agent $i$ with noise $\bm{v}_{ij,k}\sim\bm{\mathcal{N}}(\bm{0}, \bm{R}_{ij})$ under nominal conditions. The vector $\hat{\bm{z}}_{ij,k}^{\text{geo}}$ denotes the geometric estimation of the edge state for $(i,j)\in{\mathcal{E}\setminus\mathcal{E}_k}$, i.e., when the measurement is missing at time instance $k$. The covariance $\tilde{\bm{R}}_{ij,k}$ for the equivalent noise $\tilde{\bm{v}}_{ij,k}$ in (\ref{equ: state space model}) is 
\begin{equation}\label{equ: switching cov}
\tilde{\bm{R}}_{ij,k} = 
\begin{cases}
\bm{R}_{ij},&\text{if } (i,j)\in\mathcal{E}_k \\
\bm{R}^{\text{geo}}_{ij,k} + \psi_{i,k}\bm{I}_D, &\text{if } (i,j)\in{\mathcal{E}\setminus\mathcal{E}_k}
\end{cases},
\end{equation} 
where $\bm{R}^{\text{geo}}_{ij,k}$ depends on the estimator, and $\psi_{i,k}$ is an indicator function that penalizes the estimator covariance based on the confidence, which is discussed later.

Given the relative state-space model (\ref{equ: state space model}), we aim to propose adaptive filters to estimate the edge state from noisy and incomplete measurements. The estimated state could be subsequently used by the local controllers for position updates. In the following section, we present the details of the geometric estimators and discuss their performance.

\section{Geometry-Aware Estimators}\label{sec: est design}
In this section, we propose a geometric estimator and discuss the theoretical limitations and extensions. We later fuse it into a Kalman filter framework as our proposed geometric solution for $\hat{\bm{z}}_{ij,k}^{\text{geo}}$ introduced in (\ref{equ: switching observations def}). 

\subsection{Relative Affine Localization}\label{sec: RAL}
Observe that the configuration space is determined by a lower-dimensional parameter space defined by the geometric parameters $\bm{\Theta}_k^*$ and $\bm{t}^*_k$, as evident from equation (\ref{equ: target config}). As such, missing configurations can be estimated by first reconstructing this parameter space. This is a geometrical way of localizing (relative) positions, which we refer to as Relative Affine Localization (RAL). This technique assumes sufficient formation convergence as in Definition \ref{def: formation convergence} such that $\bm{z}_{i,k}$ is sufficiently close to $\bm{z}_{i,k}^*$ for $i\in\mathcal{V}$.


\begin{figure}[t]
	\centering
	
	\subfloat[]{\raisebox{0ex}
		{\includegraphics[width=0.13\textwidth]{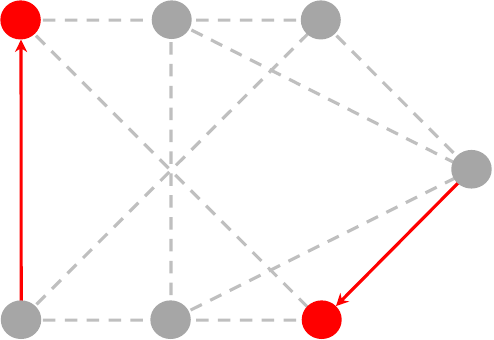}}%
	}
	\hspace{1ex}
	\subfloat[]{\raisebox{0ex}
		{\includegraphics[width=0.13\textwidth]{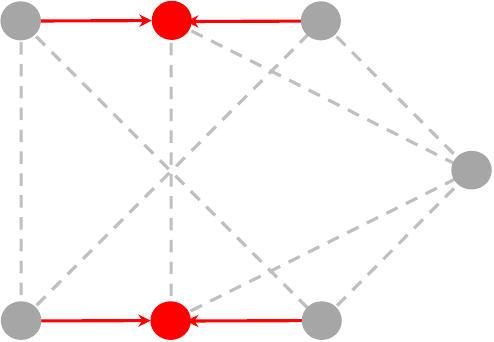}}%
	} 
	\hspace{1ex}
	\subfloat[]{\raisebox{0ex}
		{\includegraphics[width=0.13\textwidth]{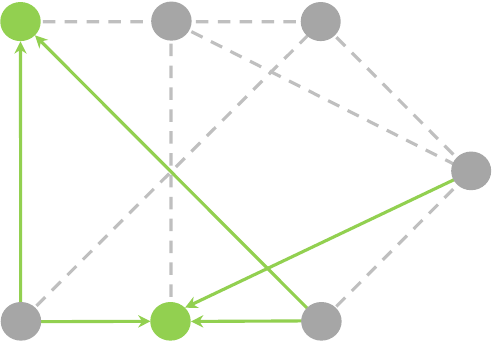}}%
	}
	
	\caption{A few examples to explain geometric feasibility of RAL in $\mathbb{R}^2$, where the colored nodes represent the agents of interest. (a) infeasible due to $N_{i,k}< D$ locally. (b) infeasible due to collinear neighbors such that $\bm{H}_{i,k}$ is not full-rank. (c) feasible.}
	\label{fig: geometry rel pos}
\end{figure}

The local form of the geometry constraint (\ref{equ: target config}) is $\bm{z}^*_{i,k} = \bm{\Theta}^*_k\bm{p}_{i} + \bm{t}^*_k$, $\forall i\in\mathcal{V}$ and observe that
\begin{equation}
    \bm{z}_{ij,k}^* = \bm{z}_{i,k}^* - \bm{z}_{j,k}^*
    = \bm{\Theta}^*_k\bm{p}_{i} + \bm{t}^*_k - (\bm{\Theta}^*_k\bm{p}_{j} + \bm{t}^*_k)
    = \bm{\Theta}^*_k\bm{p}_{ij},
    \end{equation} for $i\in\mathcal{V}, j\in\mathcal{N}_i$, where $\bm{p}_{ij} \triangleq \bm{p}_{i}-\bm{p}_{j}$ comes from the known nominal configuration. Thus $\bm{z}_{ij,k}$ is linear w.r.t. $\bm{\Theta}_k^*$, and the observations (\ref{equ: obs model}) can be extend to $\bm{y}_{ij,k} =\bm{z}_{ij,k} + \bm{v}_{ij,k} = \bm{\Theta}^*_k\bm{p}_{ij}+\bm{v}_{ij,k}$ under the assumption of sufficient formation convergence. If for each node $i\in\mathcal{V}_k$, we collect all the observations $\bm{y}_{ij,k}, \forall j\in\mathcal{N}_{i,k}$ column-wise in a matrix $\bm{Y}_{i,k}\in\mathbb{R}^{D\times N_{i,k}}$, we have a local aggregated set of equations
\begin{equation}\label{equ: obs model mat}
    \bm{Y}_{i,k} = \bm{\Theta}_k^*\bm{H}_{i,k} + \bm{V}_{i,k},
\end{equation}
where $\bm{H}_{i,k}=\bm{P}\bm{B}_{i,k}\in\mathbb{R}^{D\times N_{i,k}}$, $\bm{B}_{i,k}$  is the incidence block of agent $i$ from the functional graph defined in (\ref{equ: def inc mat}), and $\bm{V}_{i,k}\in\mathbb{R}^{D\times N_{i,k}}$ collects the corresponding noise $\bm{v}_{ij,k}$ in columns. The geometry parameters $\bm{\Theta}_k^*$ can then be estimated through a least-square formulation
\begin{equation}\label{equ: ATP estimator}
\hat{\bm{\Theta}}_{i,k}^{\text{ral}}
=\argmin_{\bm{\Theta}_k^*} 
\left \| \bm{\Theta}_k^*\bm{H}_{i,k}-\bm{Y}_{i,k}\right \|_{\mathrm{F}}^2
= \bm{Y}_{i,k}\bm{H}_{i,k}^\dagger,
\end{equation} 
where $\bm{H}_{i,k}^\dagger = \bm{H}_{i,k}\tp(\bm{H}_{i,k}\bm{H}_{i,k}\tp)^{-1}$. The reconstruction of the unobservable edge $\hat{\bm{z}}_{ij,k}^{\text{ral}}, \forall j\in{\mathcal{N}_i\setminus\mathcal{N}_{i,k}}$ can be subsequently obtained by 
\begin{equation}\label{equ: geometric estimator}
    \hat{\bm{z}}_{ij,k}^{\text{ral}} = \hat{\bm{\Theta}}_{i,k}^{\text{ral}}\bm{p}_{ij} = \bm{Y}_{i,k}\bm{H}_{i,k}^\dagger\bm{p}_{ij},
\end{equation}
with a covariance structure
\begin{equation}\label{equ: cov of geo est}
    \bm{R}^{\text{ral}}_{ij,k} = \qty(\bm{p}_{ij}\tp{\bm{H}_{i,k}^{\dagger\top}}\otimes\bm{I}_D)(\bm{I}_{N_{i,k}}\otimes \bm{R}_{ij})\qty(\bm{p}_{ij}\tp{\bm{H}_{i,k}^{\dagger\top}}\otimes\bm{I}_D)\tp.
\end{equation}

\noindent This is the RAL solution and the derivation of (\ref{equ: cov of geo est}) is shown in Appendix \ref{sec: Derivation of geo cov}, and a unique solution using (\ref{equ: geometric estimator}) is guaranteed if $\bm{H}_{i,k}$ is of full row rank, which translates to some geometric conditions which we refer to as \textit{geometric feasibility}, and define as follows.

\begin{figure*}[!t]
	\centering%
	\includegraphics[width=0.66\linewidth]{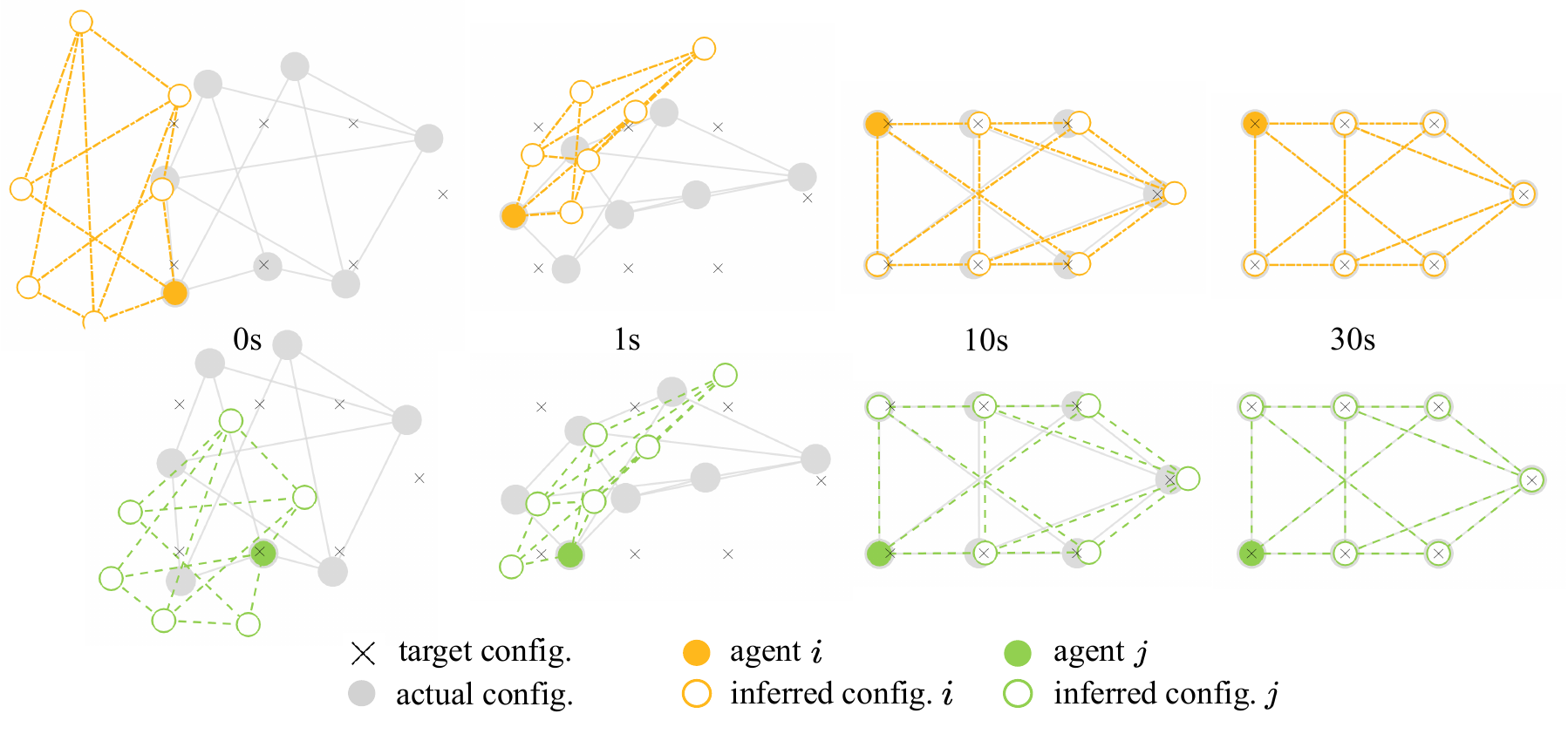}%
	\caption{An illustration of the convergence indicator at discrete time intervals of $0, 1, 10$ and $30$ seconds. While the system has not sufficiently converged, e.g., at $1$s, large discrepancies between $\hat{\bm{\Theta}}_{i,k}^\text{ral}$ and $\hat{\bm{\Theta}}_{j,k}^\text{ral}$ lead to different locally perceived formations, i.e., the inferred configurations $\hat{\bm{\Theta}}_{i,k}^\text{ral}\bm{P}$ and $\hat{\bm{\Theta}}_{j,k}^\text{ral}\bm{P}$. As the system converges to the target configuration, these discrepancies decrease until the agents reach a consensus on the perceived formation e.g., at $30$s in this illustration. This implies that the global convergence can be approximated by gauging the difference between local and neighborhood estimations of $\hat{\bm{\Theta}}_{i,k}^\text{ral}$, i.e., CI (\ref{equ: ci}).}%
	\label{fig: ci}%
\end{figure*}

\begin{definitionx}[\textit{Geometric feasibility of RAL}]\label{def: geo feas RAL}
    Relative affine localization is geometrically feasible if the estimator (\ref{equ: geometric estimator}) has a unique solution for agent $i$, when $N_{i,k}\geq D$, i.e., $\bm{Y}_{i,k}$ contains observations of at least $D$ neighboring agents that are not collinear in $\mathbb{R}^2$ or coplanar in $\mathbb{R}^3$ in the nominal configuration, to ensure a full-rank $\bm{H}_{i,k}$. A few examples in $\mathbb{R}^2$ are illustrated in Fig. \ref{fig: geometry rel pos}.
\end{definitionx} 

Although RAL applies in most cases, we are still interested in providing viable solutions in more extreme cases, i.e., for some time instance $k$, geometric feasibility is not satisfied. Since the leaders in the formation determine the maneuvering pattern, the formation can present special cases of affine transformation, such as scaling and rotation, etc., in a controlled manner. In these cases, $\bm{\Theta}_k^*$ is structured, which we can use as constraints for estimation (\ref{equ: geometric estimator}) for a further reduction of the required number of observations. We give a broad overview of this approach in Appendix \ref{sec: constrained RAL}. Another more general solution, which we provide in the following, is to allow agents to share their estimates of the geometry parameters using a consensus filter.

\subsection{Consensus Filtering}\label{sec: conRAL}
In scenarios where geometric feasibility is not met, agents can estimate $\bm{\Theta}_k^*$ using a consensus that exploits neighborhood communication. Due to the time-varying nature of the parameter of interest, i.e., $\bm{\Theta}_k^*$, we propose the use of the following dynamic consensus filter \cite{Saber2005ConsensusFilter}, which has the following updates
\begin{subequations}\label{equ: con filter}
\begin{align}
\Delta\hat{\bm{\Theta}}_{i,k} &= \sum_{j\in\mathcal{N}_{i,k}}(\hat{\bm{\Theta}}_{j,k}-\hat{\bm{\Theta}}_{i,k}) + \sum_{j\in\mathcal{J}_{i,k}}(\hat{\bm{\Theta}}_{j,k}^{\text{ral}}-\hat{\bm{\Theta}}_{i,k})\label{equ: con filter 1},\\
\hat{\bm{\Theta}}_{i,k+1} &= \hat{\bm{\Theta}}_{i,k} + \epsilon\Delta\hat{\bm{\Theta}}_{i,k},
\end{align}
\end{subequations}
where $\hat{\bm{\Theta}}_{i,k}$ are the parameters of interest for agent $i$, RAL estimates $\hat{\bm{\Theta}}_{j,k}^{\text{ral}}$ from (\ref{equ: ATP estimator}) can be considered as external inputs that contains the information of the underlying time-varying signal, and $\epsilon$ is a small constant multiplied by the increment $\Delta\hat{\bm{\Theta}}_{i,k}$. The set $\mathcal{J}_{i,k}$ contains both the neighbors and the agent $i$, i.e., $\mathcal{J}_{i,k} = \mathcal{N}_{i,k}\cup\qty{i}$. The time-varying set of neighbors $\mathcal{N}_{i,k}$ implies an intermittent communication pattern, and the convergence analysis of such behavior can be found in e.g., \cite{wen2022consensus}.


Given this filtered estimate $\hat{\bm{\Theta}}_{i,k}$, the final geometric estimation is $\hat{\bm{z}}_{ij,k}^{\text{geo}} = \hat{\bm{\Theta}}_{i,k}\bm{p}_{ij}$, which should also be the average of the RAL estimates (\ref{equ: geometric estimator}) across the nodes. The covariance of this estimate is also reduced by approximately a factor of $N$ from the (\ref{equ: cov of geo est}) 
\begin{equation}\label{equ: con filt cov}
    \bm{R}^{\text{geo}}_{ij,k} = \bm{R}^{\text{ral}}_{ij,k}/N^2,
\end{equation}
assuming no significant node failures in the network.

\subsection{Local Convergence Indicator}
Recall from Section \ref{sec: est design} that the RAL estimator assumes sufficient formation convergence. When this is not satisfied, $\hat{\bm{z}}_{ij,k}^{\text{geo}}$ in observation model (\ref{equ: switching observations def}) is subject to large biases that can be destructive to the algorithm. This is the reason we introduce the adaptive covariance matrix $\bm{R}^{\text{geo}}_{ij,k} + \psi_{i,k}\bm{I}$ in (\ref{equ: switching cov}) with a positive local convergence indicator (CI) $\psi_{i,k}$. An ideal indicator function would be the tracking error (\ref{equ: def tracking error}), which directly compares the difference between the agents' real and target positions. However, this is a global function that is locally intractable since the target configuration $\bm{Z}_k^*$ is not known to the follower agents. Therefore, we propose the following CI  
\begin{equation}\label{equ: ci}
    \psi_{i,k} = \frac{1}{N_{i,k}}\sum_{j\in\mathcal{N}_{i,k}} \norm{\hat{\bm{\Theta}}_{i,k}^\text{ral}-\hat{\bm{\Theta}}_{j,k}^\text{ral}}^2_{\mathrm{F}},
\end{equation} 
where $\hat{\bm{\Theta}}_{i,k}^\text{ral}$ is the local estimate of the geometry parameters, and $\hat{\bm{\Theta}}_{j,k}^\text{ral}$ is accessible through neighborhood communication. In contrast to (\ref{equ: def tracking error}), our proposed CI (\ref{equ: ci}) requires no global information but exhibits a similar trend as the tracking error $\delta_k$, which is suitable as the adaptive penalty of the covariance. An intuitive graphical explanation of this CI is shown in Fig. \ref{fig: ci}, and we give some formal mathematical guarantees in the following sections.

\subsection{Geometry-Aware Relative Kalman Filter}

We now reexamine the edge state-space model (\ref{equ: state space model}) and propose our edge-state tracking algorithm. Disregarding the geometric estimator $\hat{\bm{z}}_{ij,k}^{\text{geo}}$ in (\ref{equ: switching observations def}), a Kalman filter with intermittent observations can be derived from the tracking model (\ref{equ: state space model}), where the predictions based on the dynamical model are run for all time steps but the correction steps are executed only when direct edge observations $\bm{y}_{ij,k}$ are available. We name this modified Kalman filter as \textit{Relative Kalman Filter} (RKF) (\cite{Li2023GARKF} Section III B), as we focus on the edges that denote the relative state space. If the geometric estimator $\hat{\bm{z}}_{ij,k}^{\text{geo}}$ in (\ref{equ: switching observations def}) are provided, we use them as an alternative source of observations to extend the RKF framework with extra correction steps, which is coined \textit{Geometry-Aware Relative Kalman Filter} (GA-RKF), which is outlined in Algorithm \ref{alg: GARKF}.

\begin{algorithm}[t]
    \caption{\textit{GA-RKF}, for agent $i\in\mathcal{V}_k$ at time instant $k$}\label{alg: GARKF}
    \begin{algorithmic}[1]
        \State \textbf{Input}: Dimensions $D$, Number of neighbors $N_{i,k}$ and Measurements $\bm{Y}_{i,k} \in \mathbb{R}^{D \times N_{i,k}}$  (\ref{equ: obs model mat}) 
        \State \textbf{Output}: Estimates $\hat{\bm{z}}_{ij,k}$ for edge $(i,j)\in\mathcal{E}$
        \State Construct $\bm{H}_{i,k}$ corresponding to $\bm{Y}_{i,k}$

        \If{$N_{i,k} \ge D$ and $\bm{H}_{i,k}$ is full rank} 
        \State Compute $\hat{\bm{\Theta}}_{i,k}^\text{ral}$ from (\ref{equ: ATP estimator})
        \Else 
        \State Set $\hat{\bm{\Theta}}_{i,k}^\text{ral}=\bm{0}$
        \EndIf
        \State Exchange $\hat{\bm{\Theta}}_{i,k-1}$ and $\hat{\bm{\Theta}}_{i,k}^\text{ral}$ with neighbors $j\in\mathcal{N}_{i,k}$
        \State Compute convergence indicator $\psi_{i,k}$ from (\ref{equ: ci}) 
        \State Compute $\hat{\bm{\Theta}}_{i,k}$ from (\ref{equ: con filter})\Comment{Consensus filtering}
        \For{$j\in\mathcal{N}_i$}\Comment{Edge-state tracking}
            \If{$(i,j)\notin\mathcal{E}_{k}$}
                \State Compute $\hat{\bm{z}}^{\text{geo}}_{ij,k}=\hat{\bm{\Theta}}_{i,k}\bm{p}_{ij}$ 
                \State Compute $\bm{R}^{\text{geo}}_{ij,k}+\psi_{i,k}\bm{I}_D$ from (\ref{equ: cov of geo est}) and (\ref{equ: con filt cov})
            \EndIf

		\State $\bm{K}_{ij,k}=\bm{\Lambda}_{ij,k-1}\bm{G}\tp(\tilde{\bm{R}}_{ij,k}+ \bm{G}\bm{\Lambda}_{ij,k-1}\bm{G}\tp)^{-1}$
		
		\State $\hat{\bm{\gamma}}_{ij,k}=\hat{\bm{\gamma}}_{ij,k-1}+\bm{K}_{ij,k}(\tilde{\bm{y}}_{ij,k}-\bm{G}\hat{\bm{\gamma}}_{ij,k-1})$

		\State $\bm{\Lambda}_{ij,k} = (\bm{I}-\bm{K}_{ij,k}\bm{G})\bm{\Lambda}_{ij,k-1}$\Comment{Kalman iterations}

		\State $\hat{\bm{z}}_{ij,k} = \bm{G}\hat{\bm{\gamma}}_{ij,k}$
        \EndFor
		
    \end{algorithmic}
\end{algorithm}

\section{CONVERGENCE AND OPTIMALITY}\label{sec: cvg and opt}
In this section, we give insights into the conditions of convergence and optimality of the proposed GA-RKF algorithm. We also discuss the applicability and guarantees in practical scenarios. As is known, the convergence of a Kalman filter depends on the properties of the dynamic and observation matrices in the state-space model and the noise covariance. In our edge tracking case, the baseline framework is a constant acceleration model with intermittent observations (the RKF), which is well understood in the literature. We start our analysis with the adaptive coefficient $\psi_{i,k}$ in (\ref{equ: switching cov}), i.e., convergence indicator (\ref{equ: ci}).

\subsection{Convergence Indicator}
Recollect that our primary objective in introducing the convergence indicator (CI) $\psi_{i,k}$ in (\ref{equ: ci}) is to penalize the covariance if sufficient formation convergence in Definition \ref{def: formation convergence} is not satisfied. We show that $\psi_{i,k}$ qualifies as it converges the same way as the tracking error $\delta_k$ (\ref{equ: def tracking error}).

\begin{lemmax}[\textit{Upper bound of CI}]\label{lmm: upper bound of ci}
In the noiseless case, i.e., $\bm{v}_{ij,k}=0\ \forall(i,j)\in\mathcal{E}_k$, the local convergence indicator $\psi_{i,k}$ (\ref{equ: ci}) is upper bounded by the tracking error $\delta_k$ (\ref{equ: def tracking error}) as $\psi_{i,k}\leq c_{i,k}\delta_k$, where $c_{i,k}\geq0$ is a time-varying scalar given by \begin{equation}
    c_{i,k} =  \frac{N}{N_{i,k}}\sum_{j\in\mathcal{N}_{i,k}}\norm{\bm{B}_{i,k}\bm{H}^\dagger_{i,k}-\bm{B}_{j,k}\bm{H}_{j,k}^\dagger}^2_{\mathrm{F}},
\end{equation}
with known $\bm{B}_{i,k}$ and $\bm{H}^\dagger_{i,k}$ from RAL solutions (\ref{equ: geometric estimator}).
\end{lemmax}
\begin{proof}See Appendix \ref{sec: proof upb CI} \end{proof}

\noindent Lemma \ref{lmm: upper bound of ci} proves that if the tracking error $\delta_k$ follows a certain trend under standard control laws, i.e., asymptotically and exponentially converges to zero, then the local indicator $\psi_{i,k}$ will also converge similarly up to a known scalar. In the noisy case, it is intuitive that similar conclusions hold, which is formalized in the following theorem.

\begin{theoremx}[\textit{Upper bound of CI under observation noises}]\label{thm: ub under noise}
Under observation model (\ref{equ: obs model mat}), the expectation of the local convergence indicator $\mathbb{E}[\psi_{i,k}]$ is upper bounded by the expectation of the tracking error $\mathbb{E}[\delta_k]$ up to a known scaling and an offset, i.e, $\mathbb{E}[\psi_{i,k}]\leq c_{i,k}\mathbb{E}[\delta_k]+b_{i,k}$, where
\begin{subequations}\label{equ: CI scaler and offset}
    \begin{align}
        c_{i,k}&=\frac{N}{N_{i,k}}\sum_{j\in\mathcal{N}_{i,k}}\norm{\bm{B}_{i,k}\bm{H}_{i,k}^\dagger-\bm{B}_{j,k}\bm{H}_{j,k}^\dagger}^2\Fro ,\\
        b_{i,k} &= \frac{\tr(\bm{R}_{ij})}{N_{i,k}}\sum_{j\in\mathcal{N}_{i,k}}\qty(\norm{ \bm{H}_{i,k}^\dagger}^2\Fro+ \norm{\bm{H}_{j,k}^\dagger}^2\Fro).
    \end{align}
\end{subequations}
\end{theoremx}
\begin{proof}
    See Appendix \ref{sec: proof upb CI}.
    \hfill
\end{proof}

\noindent In the presence of noise, a biased term related to the noise covariance appears in the inequality. In this case, we claim that if the noise energy is finite and the tracking error converges to zero with a small margin of error, CI also converges to a finite value.

\begin{remarkx}[\textit{Sufficiency for CI convergence}]
    The convergence of the tracking error (\ref{equ: def tracking error}) is a sufficient but not a necessary condition for the convergence of the CI (\ref{equ: ci}), i.e., if $\delta_k\to 0$, then $\psi_{i,k} \to 0$ as $k\rightarrow\infty$.
\end{remarkx}
\noindent The sufficiency of CI convergence is obvious given the upper bounds in Lemma \ref{lmm: upper bound of ci} and Theorem \ref{thm: ub under noise}, but the necessity is generally not true. Note that the definition of CI (\ref{equ: ci}) only requires agreement of geometry parameters with the neighboring agents, i.e., only shape convergence of the formation is considered, but not the translations. For instance, if the formation converges to the target configuration up to a large translation, indicators $\psi_{i,k}$ for all $i$ will be zero, but the tracking error will be large. In practice, the setting of leaders will ensure convergence to the target configuration, and $\psi_{i,k}$ can be used to benchmark global convergence. 

\subsection{Convergence of GA-RKF}
In this section, we aim to give insights into the convergence of Algorithm \ref{alg: GARKF}. We begin by analyzing the baseline version of the RKF algorithm, which is a standard Kalman filter with intermittent observations \cite{Simopoli2004KFIO} without a geometric estimator. We model the availability of the edge observation as an i.i.d. Bernoulli process with parameter $0\leq\lambda_{ij}\leq1$. Here, $\lambda_{ij}=1$ indicates that all observations are available at all time instances $k$ for edge $(i,j)\in\mathcal{E}$, and $\lambda_{ij}=0$ refers to absence of any observations i.e., the edge is permanently lost from the graph. We emphasize the case when $\lambda_{ij}=0$, as this relates to the node failures scenario (see Section III \ref{sec: scenario description}). 

If the geometric estimation is not considered, the edge state-space model (\ref{equ: state space model}) is simplified to 
\begin{subequations}\label{equ: RKF ss model}
\begin{align}
    \bm{\gamma}_{ij,k+1}&=\bm{F}\bm{\gamma}_{ij,k}+\bm{w}_{ij,k},\\
    \bm{y}_{ij,k} &= \bm{G}\bm{\gamma}_{ij,k}+\bm{v}_{ij,k},
\end{align}
\end{subequations}
with $\bm{v}_{ij,k}\sim\bm{\mathcal{N}}(\bm{0}, \bm{R}_{ij})$ and $\bm{w}_{ij,k}\sim\bm{\mathcal{N}}(\bm{0}, \bm{Q}_{ij})$, and the observation availability is controlled by $\lambda_{ij}\ \forall(i,j)\in\mathcal{E}$ . Matrices $\bm{F}$ and $\bm{G}$ are the same as given for model (\ref{equ: state space model}). The following lemma states the convergence of RKF from model (\ref{equ: RKF ss model}).
\begin{lemmax}[\textit{Convergence of RKF}]\label{lmm: RKF convergence}
    If $(\bm{F},\bm{Q}_{ij}^{1/2})$ is controllable, $(\bm{F},\bm{G})$ is detectable, and $\lambda_{ij}>0$, then the estimation error covariance for RKF is bounded.
\end{lemmax}
\begin{proof}
    Let $\lambda_c$ be a known critical threshold, then we know from \cite{Simopoli2004KFIO} that if $\lambda_{ij}>\lambda_c$, then the Kalman filter with intermittent observations leads to a bounded error covariance. The proof of controllability and detectability of (\ref{equ: RKF ss model}) is straightforward and thus omitted. The critical value $\lambda_c$ is given by $\lambda_c=1-1/\alpha^2$ where $\alpha$ is the maximum of the absolute eigenvalues of $\bm{F}$, which is $1$. As such, $\lambda_{ij}>\lambda_c=0$ is needed for the convergence.
\end{proof}

\begin{remarkx}[\textit{Applicability of RKF}]
    From Lemma \ref{lmm: RKF convergence}, we observe that the critical threshold  $\lambda_c$ is zero, indicating that the RKF can achieve convergence with a minimal average number of observations, due to favorable properties of the constant acceleration model. However, RKF diverges once the edge is completely lost, i.e., $\lambda_{ij}=0$, when certain nodes are missing from the graph.
\end{remarkx}

In GA-RKF, geometric estimations provide extra edge observations, promoting the practical edge availability, especially when $\lambda_{ij}=0$. The key is to ensure a finite observation covariance $\bm{R}^{\text{geo}}_{ij,k} + \psi_{i,k}\bm{I}_D$ in model (\ref{equ: switching cov}) for the geometric estimator to qualify as an effective observation. Next, we present a general claim of the convergence of GA-RKF with an informal proof.

\begin{propositionx}[\textit{Convergence of GA-RKF}]
    At time instance $k$, GA-RKF on all edges $(i,j)\in\mathcal{E}$ admits a finite estimation covariance for $0\leq\lambda_{ij}\leq1$, if geometric feasibility in Definition \ref{def: geo feas RAL} is satisfied for agent $i$ and the tracking error $\delta_k$ is bounded.
\end{propositionx}
\begin{proof}    
    If geometric feasibility is satisfied, then $\bm{H}_{i,k}^\dagger$ in (\ref{equ: cov of geo est}) is finite. According to Theorem \ref{thm: ub under noise}, CI $\psi_{i,k}$ is also bounded if the tracking error $\delta_k$ is bounded. As such, the adaptive covariance $\bm{R}^{\text{geo}}_{ij,k} + \psi_{i,k}\bm{I}_D$ is also bounded across time to qualify as extra observations. Subsequently, the equivalent edge availability $\bar{\lambda}_{ij}$ under GA-RKF is guaranteed to be greater than the critical value shown in Lemma \ref{lmm: RKF convergence}, i.e., $\bar{\lambda}_{ij}>\lambda_c=0$ for the GA-RKF to converge.
\end{proof}

\begin{remarkx}[\textit{Applicability of GA-RKF}]
    In the scenario that geometric feasibility in Definition \ref{def: geo feas RAL} is not satisfied for some agents, GA-RKF can achieve convergence since we implement the consensus filtering (\ref{equ: con filter}) to ensure estimates from immediate neighbors are available. Hence, GA-RKF can address the $\lambda_{ij}=0$ case as compared to the RKF.
\end{remarkx}

We have given insights into the convergence property of the proposed GA-RKF algorithm, which is widely applicable owing to the geometric estimator. On the other hand, these extra observations can potentially contribute to an improved estimation performance compared to RKF, which we elaborate next by comparing the estimation bounds.

\begin{figure*}[t]
	\centering
         \subfloat[Nominal graph]{\raisebox{7ex}
    		{\includegraphics[width=0.13\textwidth]{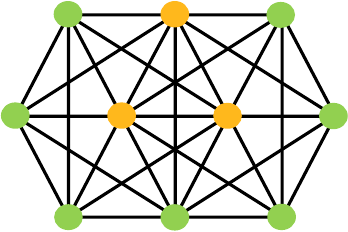}}%
    	}
        \hspace{1ex}
	\subfloat[Target trajectory used for the formation maneuver \cite{Zhao2018affine}.]{\raisebox{0ex}
		{\includegraphics[width=0.8\textwidth]{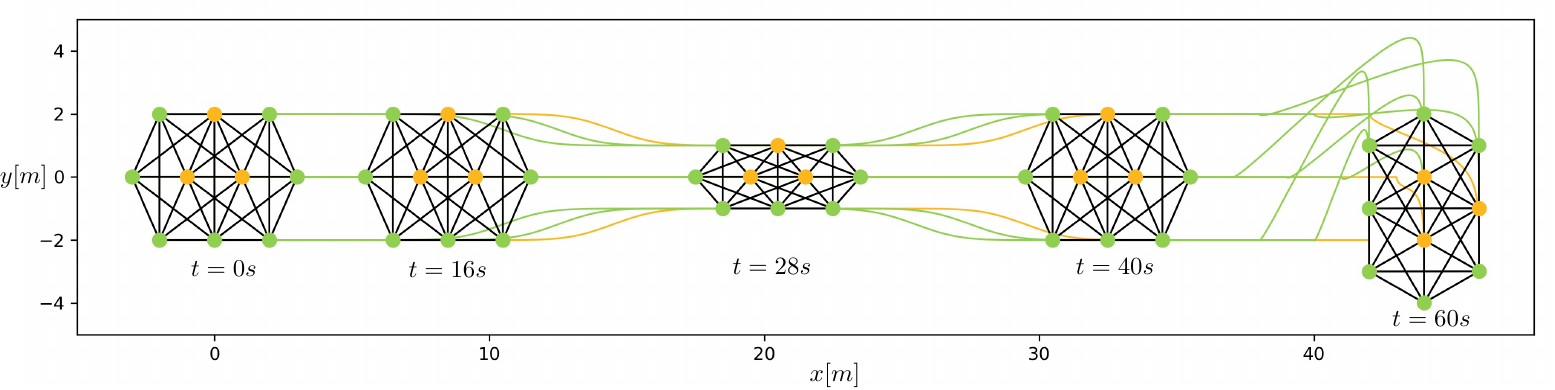}}%
	}

	\caption{The nominal graphs and the target trajectories of the formation maneuver which are used for the simulation. The nodes in orange are selected as leaders, which will not be considered in the simulation scenarios. }
	\label{fig: target traj and nominal graphs}
\end{figure*}

\subsection{Posterior Cram\'{e}r-Rao Bounds}
For a linear Gaussian state-space model, the Kalman filter is optimal i.e., the posterior covariance is minimized and reaches the posterior Cram\'{e}r-Rao bound (PCRB) \cite{tichavsky1998posterior}. The tracking model (\ref{equ: RKF ss model}) is one such case, and thus the RKF algorithm is optimal. In this section, we show that GA-RKF outperforms RKF by showing that the PCRB of the state estimate using GA-RKF is lower than that of the bound on the state estimates obtained from RKF, due to the additional geometric observations. The Fisher information matrix (FIM) for RKF under model (\ref{equ: RKF ss model}) at time instance $k$ is 
\begin{equation}\label{equ: pcrb rkf}
    \bm{J}^{\text{rkf}}_{k+1} = \beta_k\bm{G}\tp\bm{R}_{ij}^{-1}\bm{G} + \qty(\bm{Q}_{ij}+\bm{F}{\bm{J}^{\text{rkf}}_{k}}^{-1}\bm{F}\tp)^{-1},
\end{equation}
where $\beta_k$ is a binary scalar indicating the observation availability, i.e, $\beta_k=0$ when no observations are available at time $k$. Similarly, the FIM for GA-RKF under model (\ref{equ: state space model}) is
\begin{align}\label{equ: pcrb garkf}
    \bm{J}^{\text{ga-rkf}}_{k+1} &=  \beta_k\bm{G}\tp\bm{R}_{ij}^{-1}\bm{G} + (1-\beta_k)\bm{G}\tp(\bm{R}^{\text{geo}}_{ij,k}+\psi_{i,k}\bm{I}_D)^{-1}\bm{G}\notag\\
    &+\qty(\bm{Q}_{ij}+\bm{F}{\bm{J}^{\text{rkf}}_{k}}^{-1}\bm{F}\tp)^{-1},
\end{align}
where there is an additional term for the extra geometric observations. Note that FIM is additive, since our geometric estimates are independent of the direct observations. We now conclude the optimality in the following theorem.
\begin{propositionx}[\textit{Optimality of GA-RKF}]\label{prop: garkf optimality}
    The estimation covariance of GA-RKF reaches the PCRB (\ref{equ: pcrb garkf}), i.e., $\cov\qty(\hat{\bm{z}}_{ij,k})\succeq{\bm{J}^{\text{ga-rkf}}_{k}}^{-1}$ with equality, which is more optimal than PCRB for RKF (\ref{equ: pcrb rkf}), i.e., ${\bm{J}^{\text{rkf}}_{k}}^{-1}\succeq {\bm{J}^{\text{ga-rkf}}_{k}}^{-1}$.
\end{propositionx}
\begin{proof}
    It is straightforward to see that ${\bm{J}^{\text{ga-rkf}}_{k}}\succeq{\bm{J}^{\text{rkf}}_{k}}$ due to the extra positive definite term. The equality holds when $\beta_k=1$, i.e., all observations are direct observations and no geometric estimates are engaged. Then ${\bm{J}^{\text{rkf}}_{k}}^{-1}\succeq {\bm{J}^{\text{ga-rkf}}_{k}}^{-1}$ holds simply due to the Loewner ordering property \cite{horn2012matrix}. 
\end{proof}

To conclude the optimality of GA-RKF, the posterior covariance reaches the PCRB since it is still a linear Gaussian system, and its PCRB is lower than that of RKF, indicating its better optimality. The two PCRBs meet when all direct edge observations are available.

\begin{figure*}[t]
	\centering	
	\subfloat[Tracking errors $\delta_k$ under random edge losses (right) with $\lambda_{ij}=0.4$ unless indicated otherwise. $\delta_k$ is averaged across time instances $k$ (left).]{\raisebox{0.5ex}
		{\includegraphics[width=0.56\textwidth]{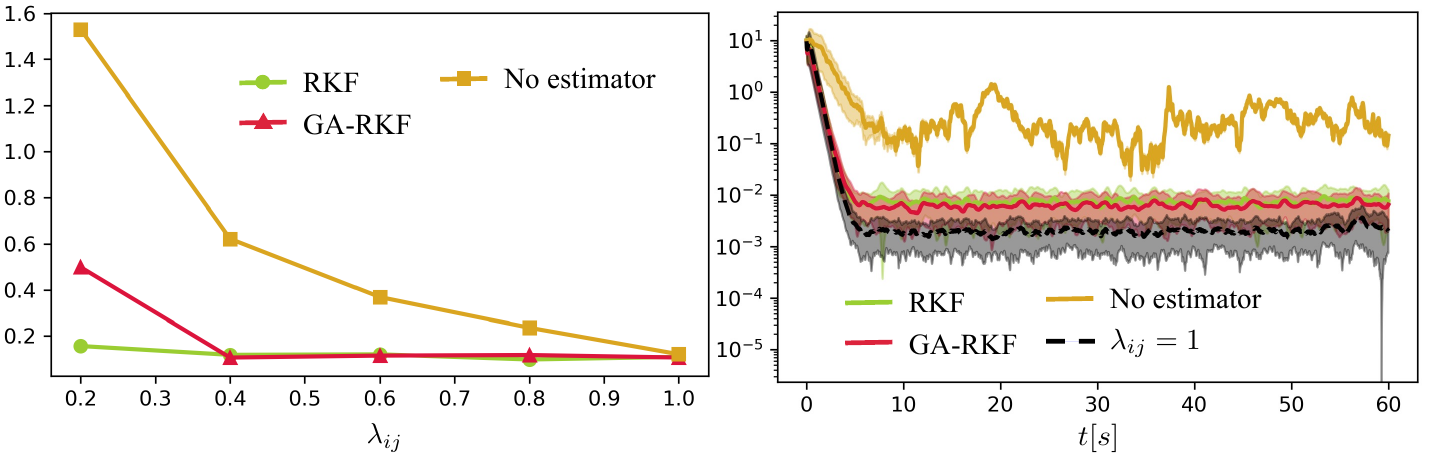}}%
	}
	\hspace{0ex}
        \subfloat[Tracking errors $\delta_k$ under node failure. $0-10s$ is executed with nominal graph in Fig. \ref{fig: target traj and nominal graphs}(a) and the remaining as shown on the left.]{\raisebox{0ex}
		{\includegraphics[width=0.42\textwidth]{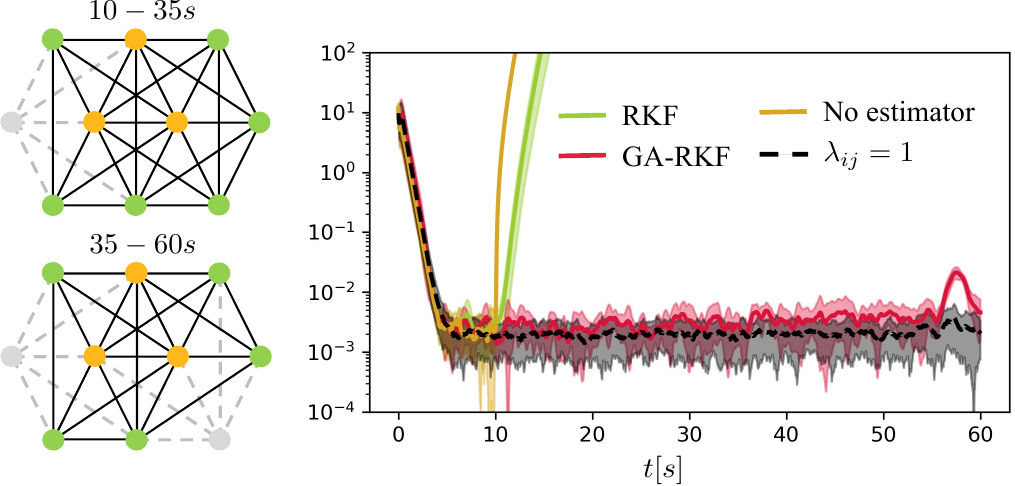}}%
	}
	
	\caption{The figure shows the tracking errors of the proposed algorithm in different practical scenarios. The presented plots are the mean over $50$ Monte Carlo runs with $\pm1$ standard deviation region.}
	\label{fig: in loop sim}
\end{figure*}

\section{SIMULATIONS}\label{sec: sim}
In this section, we present numerical validations of our proposed edge-state estimation algorithm and test it in several practical scenarios to show the enhanced robustness. We first introduce the simulation setup, where we give the nominal graphs, describe the maneuvering pattern, and various other system parameters. We then validate the optimality and convergence of the GA-RKF algorithm that is disconnected from the control loop. Finally, we show the performance of the GA-RKF algorithm when introduced into an affine formation control in some practical scenarios. The simulation code and plots are available online\footnote{https://github.com/asil-lab/zli-garkf-afc}.

\subsection{Simulation Setup}
We consider a nominal graph in $\mathbb{R}^2$ for formation maneuvers \cite{Li2023GARKF}, which is illustrated in Fig. \ref{fig: target traj and nominal graphs}(a), where there are $N=10$ nodes with $M/2 = 30$ undirected edges. The maneuvering pattern used in the simulation is shown in Fig. \ref{fig: target traj and nominal graphs}(b), where we simulate for a duration of $60$s with a discrete time step $\Delta t=0.01$s, i.e., a total of $6000$ discrete time instances. Recall from (\ref{equ: target config}) that target trajectories $\bm{Z}_k^*$ at any time instance $k \le 6000$ are determined by the underlying transformation parameters $\bm{\Theta}_k^*$ and $\bm{t}_k^*$. For the dynamical model (\ref{equ: ss dynamical model}), we opt for a constant acceleration model without loss of generality. The matrices are in standard form
\begin{align}
    \bm{F} &= \begin{bmatrix}
    1& \Delta t & \frac{1}{2}\Delta t^2 \\
    0&  1& \Delta t\\
    0& 0 &1
    \end{bmatrix}\otimes \bm{I}_D\label{equ: dynamic matrix}, \\
    \bm{Q}_{ij} &= \sigma_w^2\begin{bmatrix}
\frac{\Delta t^4}{4} & \frac{\Delta t^3}{2} & \frac{\Delta t^2}{2}\\
\frac{\Delta t^3}{2}&\Delta t^2 & \Delta t\\
\frac{\Delta t^2}{2} & \Delta t &1
\end{bmatrix}\otimes\bm{I}_D,
\end{align} 
where the variance $\sigma_w^2$ can be tuned. The noise covariance $\mathbf{R}_{ij,k}$ for the observation model (\ref{equ: obs model}) is chosen as a constant  $\sigma_v^2\mqty[1 & 0.3\\0.3&1]\ \forall k$, where $\sigma_v$ is selected as $0.1$. The step size in the consensus filter (\ref{equ: con filter}) is set to $\epsilon = 0.08$. 

\subsection{Geometry-Aware Estimators}
To validate the theoretical claims in Section \ref{sec: cvg and opt}, we first show the following results independent from the formation control, i.e., the estimates $\hat{\bm{z}}_{ij,k}$ from Algorithm \ref{alg: GARKF} are not introduced in the controller, which instead uses complete information.

\begin{figure}[t]
	\centering%
	\includegraphics[width=1\linewidth]{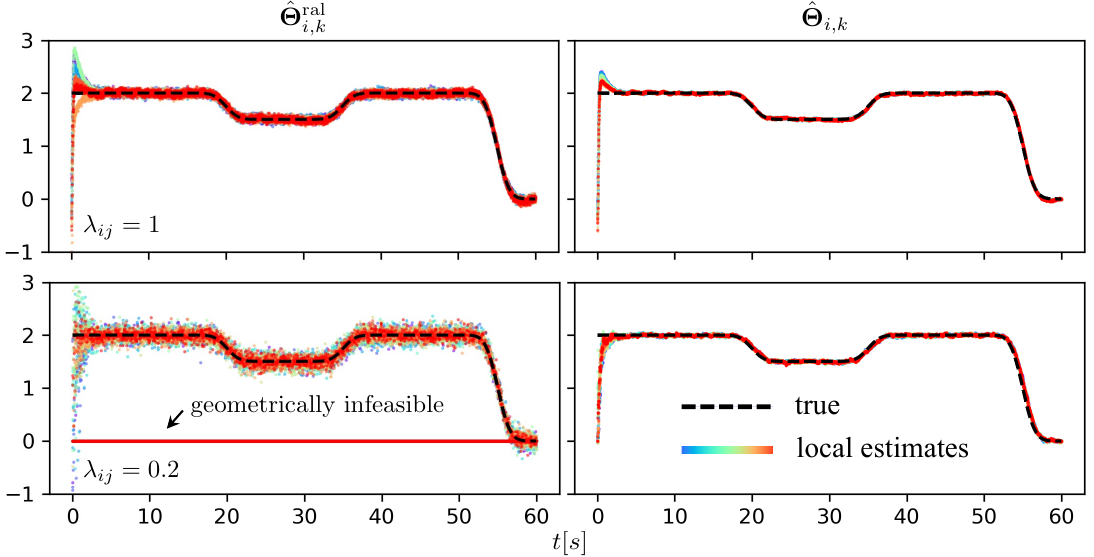}%
	\caption{Geometric estimators of $\bm{\Theta}_k^*$ from RAL (left) and consensus filtering (right). The black dashed line is the true values of $\tr(\bm{\Theta}_k^*)$ across time, and the colored dots are agents' local estimates. $\lambda_{ij}$ characterize the Bernoulli distribution chosen the same for all edges $(i,j)\in\mathcal{E}$. If $\lambda_{ij}$ is small (bottom), then a significant edge loss is present and geometric feasibility is violated more frequently, in which case $\hat{\bm{\Theta}}_{i,k}^{\text{ral}}$ is set to zero.}%
	\label{fig: est ral conral}%
\end{figure}

As discussed in Section \ref{sec: est design}\ref{sec: RAL}, the estimation of $\bm{\Theta}_k^*$ is conditioned on geometric feasibility, and the consensus filtering in Section \ref{sec: est design}\ref{sec: conRAL} is proposed to relax the condition by allowing the exchange of the estimates in the neighborhood. The estimation covariance (\ref{equ: con filt cov}) is reduced at the same time. Fig. \ref{fig: est ral conral} shows the tracking performance of the geometric estimators, where it is clearly shown that consistent estimates can be provided even when there are significant edge losses using the consensus filter. In addition, accurate tracking is achieved with reduced noise.

\begin{figure}[t]
	\centering%
	\includegraphics[width=1\linewidth]{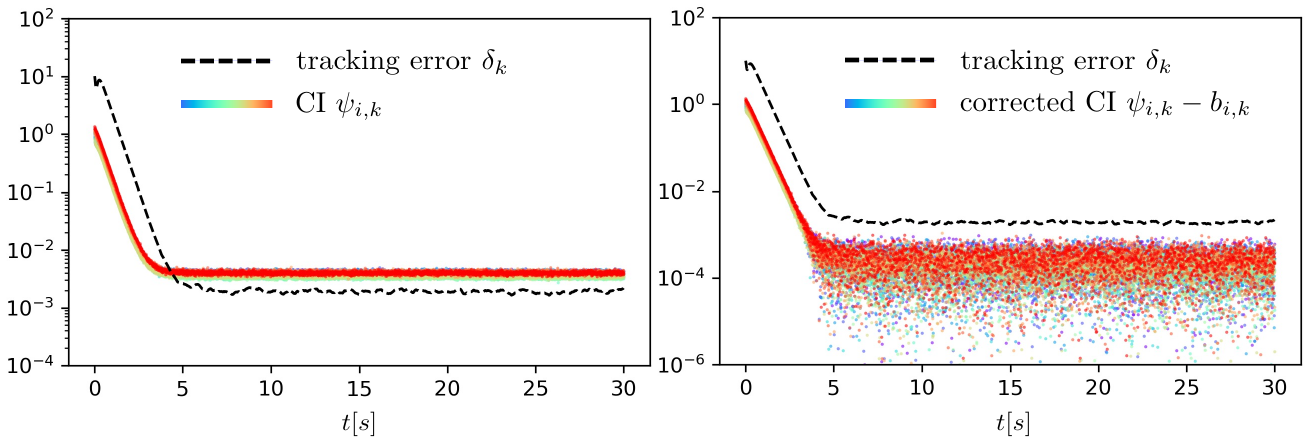}%
	\caption{The relationship between the tracking error $\delta_k$ and the convergence indicator $\psi_{i,k}$. Only the first $30s$ is simulated and the mean of $100$ Monte Carlo experiments is presented.}%
	\label{fig: sim ci}%
\end{figure}

Recall that the local convergence indicator (\ref{equ: ci}) is proposed to penalize the RAL estimates, and its boundedness and resemblance of the global tracking error (\ref{equ: def tracking error}) are claimed in Section \ref{sec: cvg and opt}, which is visualized in Fig. \ref{fig: sim ci}. As can be seen, the CI converges in similar ways to the tracking error, which indicates the formation convergence. The boundedness can be seen from the plot on the right, where the bias term in Theorem \ref{thm: ub under noise} is subtracted from the CI.

\begin{figure}[t]
	\centering%
	\includegraphics[width=1\linewidth]{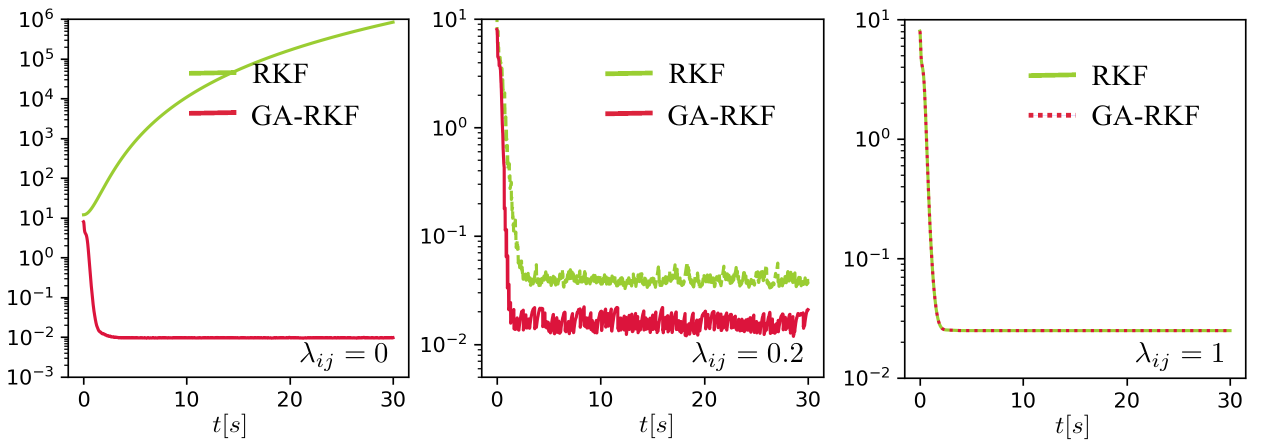}%
	\caption{The convergence of the posterior covariance of the Kalman filters. One edge in the nominal graph is selected for the visualization. $\lambda_{ij}=1$ and $\lambda_{ij}=0.2$ (middle and right) are set for all $(i,j)\in\mathcal{E}$ whereas $\lambda_{ij}=0$ (left) is only set for the selected edge and the others remain present.}%
	\label{fig: sim est cov}%
\end{figure}

We have mathematically shown the convergence and optimality of GA-RKF in previous sections, and its numerical validation is provided in Fig. \ref{fig: sim est cov}, where we compare the trace of the posterior covariance, i.e., $\tr\qty(\bm{\Lambda}_{ij,k})$ in Algorithm \ref{alg: GARKF} between GA-RKF and its root version RKF using different settings of $\lambda_{ij}$. Lemma \ref{lmm: RKF convergence} claims that RKF converges even for a small number of edge observations, but $\lambda_{ij}>0$ is still necessary to have a bounded covariance. In Fig. \ref{fig: sim est cov}, RKF diverges when $\lambda_{ij}=0$ and converges when $\lambda_{ij}=0.2$. On the other hand, GA-RKF converges even when $\lambda_{ij}=0$ due to the geometric estimates as observations. Moreover, Proposition \ref{prop: garkf optimality} shows that GA-RKF is more optimal than RKF when $\lambda_{ij}<1$ and they are equivalent when $\lambda_{ij}=1$, which is also corroborated in Fig. \ref{fig: sim est cov}.

\subsection{Practical AFC Scenarios}
Having validated the proposed theories, we now test the GA-RKF in some practical scenarios. Note that we now put the estimators in the control loop, i.e., $\hat{\bm{z}}_{ij,k}$ in Algorithm \ref{alg: GARKF} is used for the local controllers in Table \ref{tab: control laws}. We use the tracking error (\ref{equ: def tracking error}) as our main evaluation metric, and the performance of convergence and optimality of AFC are focused on.

We first set up random edge losses, which cover scenarios S1 and potentially S2 motivated in Section \ref{sec: prob form}\ref{sec: scenario description}. We simulate the tracking error across a spectrum of $\lambda_{ij}$s. Fig. \ref{fig: in loop sim}(a) shows the tracking error for the proposed algorithm, where both RKF and GA-RKF have a significant improvement on the average tracking error compared with the case where no estimators are implemented, especially under low $\lambda_{ij}$. In the particular case where $\lambda_{ij}=0.4$ shown on the right in Fig. \ref{fig: in loop sim}(a), we see that the convergence speed is maintained compared with the $\lambda_{ij}=1$ case, where all edge observations are available, which is used as a experimental bound on the tracking error.

Fig. \ref{fig: in loop sim} (b) shows another scenario where some agents (followers) are disconnected, which is motivated by S3 in Section \ref{sec: prob form}\ref{sec: scenario description}, and we expect the rest to remain stable and maintain their respective target positions. As can be seen, the "No estimator" case and RKF diverge with node failure since the underlying topology change requires a new stabilizing controller, if it exists at all, due to graph rigidity requirements. For the RKF, the predictions are outdated as soon as the relative dynamics change without any correction of observations. On the other hand, since the geometry estimates are still accurate as the node failures do not instantly corrupt the geometric pattern. The remaining nodes estimate the edge as if the failed nodes are keeping the formation, and thus could maintain their target positions. As such, we conclude that the GA-RKF algorithm is robust to this node failure case.

\section{CONCLUSIONS}\label{sec: conclusions}
In this paper, we proposed the GA-RKF algorithm to enhance the resilience of AFC to topological changes induced by observation losses. Our algorithm adaptively fuses temporal and spatial (geometrical) information using a Kalman filtering framework that results in the convergence in extreme settings, which are shown by theoretical results and numerical experiments. We have also shown the practical applicability of GA-RKF in random edge loss and permanent node failure scenarios. In our future work, we aim to strengthen several aspects of our proposed framework further. First, the discussions are limited to the edge availability and node departures w.r.t. followers. Furthermore, we would like to generalize our solutions to general directed graphs.


\appendices
\section{Derivation of the covariance (\ref{equ: cov of geo est})}\label{sec: Derivation of geo cov}
Given model (\ref{equ: obs model mat}) and estimator (\ref{equ: geometric estimator}), we have
\begin{align}
    \hat{\bm{z}}_{ij,k}^{\text{ral}} = \bm{Y}_{i,k}\bm{H}_{i,k}^\dagger\bm{p}_{ij}
    = (\bm{\Theta}_k^*\bm{H}_{i,k} + \bm{V}_{i,k})\bm{H}_{i,k}^\dagger\bm{p}_{ij},
\end{align} where $\bm{V}_{i,k}$ is the stochastic element. Applying the property of vectorization and the Kronecker product (see (520) in \cite{petersen2008matrix}), the covariance of the estimator is
\begin{subequations}
   \begin{align}
    \cov\qty(\hat{\bm{z}}_{ij,k}^{\text{ral}}) &= \cov\qty(\vec\qty(\hat{\bm{z}}_{ij,k}^{\text{ral}}))\\
    &=\cov\qty(\vec\qty(\bm{V}_{i,k}\bm{H}_{i,k}^\dagger\bm{p}_{ij}))\\
    &= \cov\qty(\qty(\bm{p}_{ij}\tp{\bm{H}_{i,k}^{\dagger\top}}\otimes\bm{I}_D)\vec\qty(\bm{V}_{i,k}))\\ 
    &= \qty(\bm{p}_{ij}\tp{\bm{H}_{i,k}^{\dagger\top}}\otimes\bm{I}_D)\qty(\bm{I}_{N_{i,k}}\otimes\bm{R}_{ij}) \notag\\
    &\qty(\bm{p}_{ij}\tp{\bm{H}_{i,k}^{\dagger\top}}\otimes\bm{I}_D)\tp,
\end{align} 
\end{subequations}
where we assume the noise vector $\bm{v}_{ij}$ across the edges are i.i.d. with covariance $\bm{R}_{ij}$.

\section{Constrained RAL}\label{sec: constrained RAL}
In this section, we give an overview of the solutions when the formation is restricted to special cases of affine transformations. An overview of these constrained problems and their solutions is shown in Table \ref{tab: cstr RAL}.

\begin{table*}[!tb]
	\caption{Solutions for (\ref{equ: ATP estimator}) with additional constraints}
	\label{tab: cstr RAL}
	\centering
	\begin{tabular}{l c c c c c}
		\toprule
		{} & General & Translation & Scaling & Rotation & Similarity\\
		\midrule
		{constraints} & - & $\bm{\Theta}_k^*=\bm{I}_D$ & $\bm{\Theta}_k^*$ is diagonal & $\bm{\Theta}_k^{*\top}\bm{\Theta}_k^* = \bm{I}_D$ & $\bm{\Theta}_k^{*\top}\bm{\Theta}_k^* = s_k^2\bm{I}_D$\\ \\

             {solutions}& (\ref{equ: ATP estimator}) & $\hat{\bm{\Theta}}_{i,k}^{\text{ral}}=\bm{I}_D$ &$\hat{\bm{\Theta}}_{i,k}^{\text{ral}} = \diag(\hat{s}_{1,k},...,\hat{s}_{D,k})$ & $\begin{aligned}  \bm{Y}_{i,k}\bm{H}_{i,k}\tp &= \bm{U}_{i,k}\bm{\Sigma}_{i,k}\bm{\Phi}_{i,k}\tp  \\
  \hat{\bm{\Theta}}_{i,k}^{\text{ral}} &= \bm{U}_{i,k}\bm{\Phi}_{i,k}\tp 
\end{aligned}$ & $\begin{aligned}
    \hat{s}_{i,k} &= \frac{1}{D}\tr({\bm{\Sigma}_{i,k}^{H}}^{-1}\bm{\Sigma}^Y_{i,k})\\
    \hat{\bm{\Theta}}_{i,k}^{\text{ral}} &= \hat{s}_{i,k}\bm{U}_{i,k}\bm{\Phi}_{i,k}\tp
    
\end{aligned}$\\
		
		 \\ minimum $N_{i,k}$ & $D$ & $0$ & $1$ & $D-1$ & $D-1$ \\
	
		\bottomrule
	\end{tabular}
\end{table*}
\textit{Translation only:} Translation is dictated by the $\bm{t}_k^*$ vector in (\ref{equ: target config}), and the shape of the geometry, represented by $\bm{\Theta}_k^*$, is not changed. As such, the estimation of $\bm{\Theta}_k^*$ can simply be set to $\hat{\bm{\Theta}}_{i,k} = \bm{I}_D$. 

\textit{Scaling only:}
Scaling of the geometry can allow the swarm to pass through narrow passages or around obstacles. For cases where only scaling is involved, the parameter matrix $\bm{\Theta}_k^*$ degenerates to
\begin{equation}\label{equ: def scaling only}
    \bm{\Theta}_k^* = \diag\qty(s_k^1,...,s_k^D),
\end{equation}
where $s^d_k\in\mathbb{R}$, for $d=1,...,D$, scales each dimension. Since the diagonality of $\bm{\Theta}_k^*$ decouples the dimensions in (\ref{equ: ATP estimator}), the problem could be decomposed into $D$ simple independent least-squares problems.

\textit{Rotation only:}
The parameter matrix $\bm{\Theta}_k^*$ in this case is an orthogonal rotation matrix with orthonormal columns, i.e., $\bm{\Theta}_k^{*\top}\bm{\Theta}_k^* = \bm{I}_D$ is constrained for (\ref{equ: ATP estimator}). This formulation is recognized as the orthogonal Procrustes problem with analytical solutions available \cite{Schnemann1966AGS}. As presented in Table \ref{tab: cstr RAL}, the solution is derived from a singular value decomposition (SVD) $\bm{Y}_{i,k}\bm{H}_{i,k}\tp = \bm{U}_{i,k}\bm{\Sigma}_{i,k}\bm{\Phi}_{i,k}\tp$ where $\bm{U}_{i,k}$ and $\bm{\Phi}_{i,k}$ contain the singular vectors and $\bm{\Sigma}_{i,k}$ with singular values.

\textit{Similarity transform:}
Similarity transform is a combination of rotation and scaling where each dimension is uniformly scaled by a non-negative $s_k$ on top of a rotation. Hence, the constraints on $\bm{\Theta}_{k}^*$ can be relaxed with a scaling ambiguity, as shown in Table \ref{tab: cstr RAL}. The rotation and scaling can be estimated independently, with the rotation estimated in the same manner as in the previous rotation-only case. For the scaling, if $\bm{H}_{i,k}$ and $\bm{Y}_{i,k}$ both have economy-sized SVD with $\bm{\Sigma}^H_{i,k}$ and $\bm{\Sigma}^Y_{i,k}$ as the respective diagonal matrices with singular values, $s_k$ can be estimated by $\hat{s}_{i,k} = \frac{1}{D}\tr({\bm{\Sigma}_{i,k}^{H}}^{-1}\bm{\Sigma}^Y_{i,k})$.

\section{Convergence Indicator (CI)}\label{sec: proof upb CI}
\subsection{Proof of Lemma \ref{lmm: upper bound of ci}}
\begin{proof}
In the noiseless case, model (\ref{equ: obs model mat}) becomes $\bm{Y}_{i,k} = \bm{Z}_{k}\bm{B}_{i,k} = \bm{\Theta}_k^*\bm{H}_{i,k}$, and the solution to (\ref{equ: ATP estimator}) becomes $\hat{\bm{\Theta}}_{i,k}^{\text{ral}} = \bm{Z}_{k}\bm{B}_{i,k}\bm{H}_{i,k}^\dagger$, where $\bm{B}_{i,k}$ is the incidence block for agent $i$ in the time-varying functional graph. Multiplying $\bm{B}_{i,k}$ on both sides of definition (\ref{equ: target config}) gives $\bm{Z}^*_k\bm{B}_{i,k} = \bm{\Theta}^*_k\bm{P}\bm{B}_{i,k} = \bm{\Theta}^*_k\bm{H}_{i,k}$, which yields in $\bm{\Theta}_k^*= \bm{Z}^*_k\bm{B}_{i,k}\bm{H}_{i,k}^\dagger$. Then (\ref{equ: ci}) can be rewritten as follows
\begin{subequations}\label{equ: dev ci}
    \begin{align}
        \psi_{i,k} &= \frac{1}{N_{i,k}}\sum_{j\in\mathcal{N}_{i,k}} \norm{\hat{\bm{\Theta}}_{i,k}^\text{ral}-\hat{\bm{\Theta}}_{j,k}^\text{ral}}^2\Fro\\
        &= \frac{1}{N_{i,k}}\sum_{j\in\mathcal{N}_{i,k}}\norm{\qty(\hat{\bm{\Theta}}_{i,k}^\text{ral}-\bm{\Theta}^*_k)-\qty(\hat{\bm{\Theta}}_{j,k}^\text{ral}-\bm{\Theta}^*_k)}^2\Fro\\
        &= \frac{1}{N_{i,k}}\sum_{j\in\mathcal{N}_{i,k}}\norm{\qty(\bm{Z}_k-\bm{Z}^*_k)\qty(\bm{B}_{i,k}\bm{H}_{i,k}^\dagger-\bm{B}_{j,k}\bm{H}_{j,k}^\dagger)}^2\Fro\label{equ: CI before inequ}\\
        & \leq \frac{1}{N_{i,k}}\sum_{j\in\mathcal{N}_{i,k}}\norm{\bm{B}_{i,k}\bm{H}_{i,k}^\dagger-\bm{B}_{j,k}\bm{H}_{j,k}^\dagger}^2\Fro\norm{\bm{Z}_k-\bm{Z}^*_k}^2\Fro\\
        &\leq \frac{N}{N_{i,k}}\sum_{j\in\mathcal{N}_{i,k}}\norm{\bm{B}_{i,k}\bm{H}_{i,k}^\dagger-\bm{B}_{j,k}\bm{H}_{j,k}^\dagger}^2\Fro \delta_k \\
        &\leq c_{i,k}\delta_k,
    \end{align}
\end{subequations} and hence proven.
\end{proof}

\subsection{Proof of Theorem \ref{thm: ub under noise}}\label{sec: proof sto CI}
\begin{proof}
Recollect from the definition of CI (\ref{equ: ci}) and use similar trick as in (\ref{equ: dev ci}),
\begin{subequations}
    \begin{align}
        \psi_{i,k} &= \frac{1}{N_{i,k}}\sum_{j\in\mathcal{N}_{i,k}} \norm{\hat{\bm{\Theta}}_{i,k}^\text{ral}-\hat{\bm{\Theta}}_{j,k}^\text{ral}}^2\Fro\\
        &= \frac{1}{N_{i,k}}\sum_{j\in\mathcal{N}_{i,k}}\norm{\qty(\hat{\bm{\Theta}}_{i,k}^\text{ral}-\bm{\Theta}^*_k)-\qty(\hat{\bm{\Theta}}_{j,k}^\text{ral}-\bm{\Theta}^*_k)}^2\Fro\\
        &= \frac{1}{N_{i,k}}\sum_{j\in\mathcal{N}_{i,k}}\normx{\qty(\bm{Z}_k-\bm{Z}^*_k)\qty(\bm{B}_{i,k}\bm{H}_{i,k}^\dagger-\bm{B}_{j,k}\bm{H}_{j,k}^\dagger)\notag\\
        &+\qty(\bm{V}_{i,k}\bm{H}_{i,k}^\dagger-\bm{V}_{j,k}\bm{H}_{j,k}^\dagger)}a^2\Fro\\
        &=\psi_{i,k}'+\psi_{i,k}''+\psi_{i,k}''',
    \end{align}
\end{subequations}
where the expressions for $\psi_{i,k}'$, $\psi_{i,k}''$ and $\psi_{i,k}'''$ are given by
\begin{subequations}\label{equ: psi primes}
\begin{align}
\psi_{i,k}' &= \frac{1}{N_{i,k}}\sum_{j\in\mathcal{N}_{i,k}}\norm{\qty(\bm{Z}_k-\bm{Z}^*_k)\qty(\bm{B}_{i,k}\bm{H}_{i,k}^\dagger-\bm{B}_{j,k}\bm{H}_{j,k}^\dagger)}^2\Fro\\
\psi_{i,k}''&=\frac{1}{N_{i,k}}\sum_{j\in\mathcal{N}_{i,k}}\norm{\bm{V}_{i,k}\bm{H}_{i,k}^\dagger-\bm{V}_{j,k}\bm{H}_{j,k}^\dagger}^2\Fro\\
\psi_{i,k}''' &=\frac{2}{N_{i,k}}\sum_{j\in\mathcal{N}_{i,k}}\tr\bigg(\qty(\bm{B}_{i,k}\bm{H}_{i,k}^\dagger-\bm{B}_{j,k}\bm{H}_{j,k}^\dagger)\tp\notag \\
&\qty(\bm{Z}_k-\bm{Z}^*_k)\tp\qty(\bm{V}_{i,k}\bm{H}_{i,k}^\dagger-\bm{V}_{j,k}\bm{H}_{j,k}^\dagger)\bigg).
\end{align}
\end{subequations}

\noindent We first define the expectation of the tracking error $\delta_k$ and the expectation of CI as 
\begin{align}
    \mathbb{E}\qty[\delta_k] &= \frac{1}{N}\mathbb{E}\qty[\norm{ \bm{Z}_k-\bm{Z}^*_k}^2\Fro], \\
    \mathbb{E}\qty[\psi_{i,k}] &=\mathbb{E}\qty[\psi_{i,k}']+\mathbb{E}\qty[\psi_{i,k}'']+\mathbb{E}\qty[\psi_{i,k}''']\label{equ: exp psi primes},
\end{align}
respectively, and we then analyze each term individually.

Since $\psi_{i,k}'$ is the same as (\ref{equ: CI before inequ}), we have 
\begin{subequations}\label{equ: psi'}
\begin{align}
    \mathbb{E}[\psi_{i,k}'] &\leq 
    \mathbb{E}\bigg[\frac{1}{N_{i,k}}\sum_{j\in\mathcal{N}_{i,k}}\norm{\bm{B}_{i,k}\bm{H}_{i,k}^\dagger-\bm{B}_{j,k}\bm{H}_{j,k}^\dagger}^2\Fro \notag \\
    &\norm{ \bm{Z}_k-\bm{Z}^*_k}^2\Fro\bigg]\\
    & = \frac{N}{N_{i,k}}\sum_{j\in\mathcal{N}_{i,k}}\norm{\bm{B}_{i,k}\bm{H}_{i,k}^\dagger-\bm{B}_{j,k}\bm{H}_{j,k}^\dagger}^2\Fro\mathbb{E}\qty[\delta_k].
\end{align}
\end{subequations}For term $\mathbb{E}[\psi_{i,k}'']$ in (\ref{equ: exp psi primes}),
\begin{subequations}\label{equ: psi''}
\begin{align}
    \mathbb{E}[\psi_{i,k}'']&=\frac{1}{N_{i,k}}\sum_{j\in\mathcal{N}_{i,k}}\tr\qty(\bm{H}_{i,k}^{\dagger\top}\mathbb{E}\qty[\bm{V}_{i,k}\tp\bm{V}_{i,k}]\bm{H}_{i,k}^\dagger)\notag\\
    &+\tr\qty(\bm{H}_{j,k}^{\dagger\top}\mathbb{E}\qty[\bm{V}_{j,k}\tp\bm{V}_{j,k}]\bm{H}_{j,k}^\dagger)\notag\\
    &-2\tr\qty(\bm{H}_{i,k}^{\dagger\top}\mathbb{E}\qty[\bm{V}_{i,k}\tp\bm{V}_{j,k}]\bm{H}_{j,k}^\dagger)\label{equ: expand trace CI}\\
    &=\frac{1}{N_{i,k}}\sum_{j\in\mathcal{N}_{i,k}}\tr\qty(\bm{R}_{ij})\tr\qty(\bm{H}_{i,k}^{\dagger\top}\bm{H}_{i,k}^\dagger)\notag\\   &+\tr\qty(\bm{R}_{ij})\tr\qty(\bm{H}_{j,k}^{\dagger\top}\bm{H}_{j,k}^\dagger) + 0 \label{equ: expanded trace CI}\\
    &= \frac{\tr\qty(\bm{R}_{ij})}{N_{i,k}}\sum_{j\in\mathcal{N}_{i,k}}\norm{\bm{H}_{i,k}^\dagger}^2\Fro+\norm{\bm{H}_{j,k}^\dagger}^2\Fro,
\end{align}    
\end{subequations} where we use Property \ref{prpt: expect inner prod noise mat} in Appendix \ref{sec: properties apdx}.  Finally, for term $\mathbb{E}[\psi_{i,k}''']$ in (\ref{equ: exp psi primes}), we have \begin{subequations}\label{equ: psi'''}
\begin{align}
    \mathbb{E}[\psi_{i,k}'''] &= \mathbb{E}\bigg[\frac{1}{N_{i,k}}\sum_{j\in\mathcal{N}_{i,k}}\tr\bigg(\qty(\bm{B}_{i,k}\bm{H}_{i,k}^\dagger-\bm{B}_{j,k}\bm{H}_{j,k}^\dagger)\tp\notag\\
    &(\bm{Z}_k-\bm{Z}^*_k)\tp\qty(\bm{V}_{i,k}\bm{H}_{i,k}^\dagger-\bm{V}_{j,k}\bm{H}_{j,k}^\dagger)\bigg)\bigg] \\
    &= \frac{1}{N_{i,k}}\sum_{j\in\mathcal{N}_{i,k}}\tr\bigg(\qty(\bm{B}_{i,k}\bm{H}_{i,k}^\dagger-\bm{B}_{j,k}\bm{H}_{j,k}^\dagger)\tp\notag\\
    &\mathbb{E}\qty[\bm{Z}_k-\bm{Z}^*_k]\tp\qty(\mathbb{E}\qty[\bm{V}_{i,k}]\bm{H}_{i,k}^\dagger-\mathbb{E}\qty[\bm{V}_{j,k}]\bm{H}_{j,k}^\dagger)\bigg) \nonumber \\
    & = 0,
\end{align}    
\end{subequations}
as the noises in matrices $\bm{V}_{i,k}$ for all $i$ and $k$ are zero-mean and uncorrelated with $\bm{Z}_k-\bm{Z}^{*}_k$. Combing the expressions (\ref{equ: psi'}), (\ref{equ: psi''}) and (\ref{equ: psi'''}), we can conclude that
\begin{align}
\mathbb{E}[\psi_{i,k}] &\leq c_{i,k}\mathbb{E}[\delta_k]+b_{i,k},
\end{align} 
where $c_{i,k}$ and $b_{i,k}$ are given by (\ref{equ: CI scaler and offset}) in Theorem \ref{thm: ub under noise}.
\end{proof}

\subsection{Expectation of Noise Matrix Inner Products}
\label{sec: properties apdx}
\begin{propertyx}\label{prpt: expect inner prod noise mat}
Given a matrix $\bm{V}=[\bm{v}_1,\bm{v}_2,...,\bm{v}_N]\in\mathbb{R}^{D\times N}$ where $\bm{v}_i\overset{\text{i.i.d}}{\sim}\bm{\mathcal{N}}(\bm{0}_D,\bm{R})$ for $i=1,...,N$, the following statement holds true
\begin{equation}
    \mathbb{E}\qty[\bm{V}\tp\bm{V}] = \tr\qty(\bm{R})\bm{I}_N.
\end{equation}
\end{propertyx}
\begin{proof}
If we use $i$ and $j$ to denote the matrix row and column index, respectively, then elements in matrix $\mathbb{E}[\bm{V}^T\bm{V}]$ are
\begin{align}
        \qty[\mathbb{E}\qty[\bm{V}\tp\bm{V}]]_{ij}=\left\{\begin{matrix} 
 \mathbb{E}\qty[\bm{v}_i\tp\bm{v}_j], & i=j\\  
  0, &\text{otherwise}
\end{matrix}\right. ,
    \end{align}
since $\bm{v}_i$ and $\bm{v}_j$ are independent for any $i=1,2,\hdots, N$ and $j\ne i$. In addition, the expectation of $\bm{v}\tp_{i}\bm{v}_{i}$ is the sum of the variance of each element in the vector i.e., $\mathbb{E}[\bm{v}\tp_{i}\bm{v}_{i}]=\text{tr}[\bm{R}]$. Moreover, since the vectors are also identically distributed, we have $\mathbb{E}[\bm{V}\tp\bm{V}]=\text{tr}[\bm{R}]\bm{I}_N$, and hence proven.
\end{proof}

\bibliographystyle{IEEEtran}
\bibliography{refs}

\end{document}